# Physical properties of transition metal hydride superconductors Mg$_2$*Tm*H$_6$ (*Tm* = Rh, Pd, Ir, Pt) by first-principles calculations

*Md Ashraful Alam[1,2], Md Abdul Hadi Shah[2,3], F. Parvin[2], S. H. Naqib[2,]\**

[1]*Department of Physics, Mawlana Bhashani Science and Technology University, Santosh, Tangail 1902, Bangladesh*
[2]*Department of Physics, University of Rajshahi, Rajshahi 6205, Bangladesh*
[3]*Department of Physics, Rajshahi University of Engineering and Technology, Rajshahi 6204, Bangladesh*
*\*Corresponding author*: Email*: salehnaqib@yahoo.com*

## Abstract

In this work, a comprehensive first-principles investigation of the structural, hydrogen storage potential, electronic, elastic, mechanical, thermophysical, superconducting, and optical properties of Mg$_2$*Tm*H$_6$ (*Tm* = Rh, Pd, Ir, Pt) hydrides is presented. All compounds are thermodynamically, mechanically, and dynamically stable, with hydrogen storage capacities ranging from 2.42 to 3.84 wt%. All behave as ductile, elastically anisotropic, and exhibit weak metallic character with strong directional bonding, reflected in the elastic hierarchy $C_{11} > C_{12} > C_{44}$ as well as the moduli trend Y > B > G. Mg$_2$PtH$_6$ possesses the largest bulk modulus, indicating the lowest compressibility. In contrast, Mg$_2$IrH$_6$ exhibits the highest Young's modulus, superior machinability, and excellent dry lubricity. Electronic structure analyses show low density of states at the Fermi level, with predicted superconducting transition temperatures of 25–44 K. Thermophysical calculations reveal their stability and efficient heat transport. Optical studies indicate high reflectivity in the infrared and visible regions, decreasing in the ultraviolet, while strong UV absorption and optical conductivity suggest potential for UV optoelectronic devices, photodetectors, and reflective coatings. These results demonstrate that Mg$_2$*Tm*H$_6$ hydrides combine favorable hydrogen storage, mechanical robustness, superconductivity, and multifunctional optical properties, making them promising candidates for energy storage, superconducting and advanced optoelectronic applications.

**Keywords:** Transition metal hydrides; Hydrogen storage capacity; DFT calculations; Elastic properties; Thermophysical properties; Superconducting transition temperature

1. **Introduction**

Hydrogen is an ideal energy carrier for reducing dependence on imported oil, particularly for countries lacking natural resources. Hydrides are compounds formed by the combination of hydrogen with metals or metalloids. Based on the nature of chemical bonding, hydrides can be classified into four categories (i) ionic hydride (ii) covalent hydride (iii) metallic hydride and (iv) van der Waals hydride [1]. Hydrides are important for various applications, such as hydrogen storage systems [2], cryo-coolers [3], catalysis [4], and the nuclear technology industry [5–9]. In addition, hydrides also exhibit high-temperature superconducting properties. Hydrides are promising materials due to their high hydrogen storage capacity, abundance of raw materials, low cost and potential superconducting properties [10–13]. Hydrides exhibit good gravimetric hydrogen storage capacity, for example $MgH_6$ (7.6 wt%) [12], $CoCuH_3$ (2.80 wt%) [14], $KCuH_3$ (2.86 wt%) [15], $NaScH_3$ (4.26 wt%) [2], $KSrH_3$ (2.33 wt%) [16], $B_{20}H_{16}$ (6.90 wt%) [17], $LaMg_2Ni$ (1.96 wt%) [18], and $Mg_{15}ScH_{32}$ (7.3 wt%) [19].

Binary metallic hydrides exhibit high temperature superconducting properties, such as $H_3S$ [20,21] ($T_C$ ~203 K at 155 GPa), $YH_9$ ($T_C$ of 253−276 K at 200 GPa) [22], $YH_6$ ($T_C$ of 251−264 K at 110 GPa) [23] and $YH_{10}$ ($T_C$ ~326 K at 250 GPa) [24], $UH_8$ ($T_C$ ~ 204 K, $T_C$ ~ 193 K at 200 GPa and under ambient condition, respectively) [25]. In addition, a large number of ternary hydrides have been predicted to possess excellent superconducting properties under pressure, including $ScCaH_8$ and $ScCaH_{12}$ ($T_C$ ~212 K and ~182 K, respectively, at 200 GPa) [26], $ScYH_6$ ($T_C$ ~32.11 K to 52.90 K in the pressure range 0–200 GPa) [27], $H_3SXe$ ($T_C$ of 89 K at 240 GPa) [28], $CaYH_{12}$ ($T_C$ ~258 K at 200 GPa) [29], $LaSH_6$ ($T_C$ ~35 K at 300 GPa) [30]. It is noteworthy that Mg-based ternary hydrides also exhibit superconducting properties under pressure, such as $MgSiH_6$ ($T_C$ ~ 63 K at 250 GPa) [31], $MgScH_6$ ($T_C$ 41 K at 100 GPa) [32], $MgVH_6$ ($T_C$ of 27.6 K at 150 GPa) [13]. In addition, Wang *et al*. [33] proposed two ternary hydrides, $Mg_2FeH_6$ and $Mg_2RuH_6$ which were predicted to possess semiconducting properties.

From the above discussion, it is evident that hydrides have attracted considerable attention from the scientific community. Typically, hydrides exhibit insulating behavior rather than superconductivity at low pressure, whereas at high pressure many superconducting hydrides become structurally unstable. One of the major challenges in the field of hydride superconductivity is achieving superconductivity at low or ambient pressure, rather than at the extremely high pressures typically required in experiments. Only a few hydrides have been proposed in the literature to exhibit high-temperature superconductivity under ambient pressure. A family of Mg-based transition-metal hydrides, $Mg_2TmH_6$ ($Tm$ = Rh, Ir, Pd, Pt), was proposed by Sanna *et al.* [34] as potential high-temperature conventional superconductors. These materials are predicted to be both structurally and thermodynamically stable and are closely related to the experimentally synthesized $Mg_2RuH_6$ [35,36]. All compounds were identified using the Alexandria Materials Database [37,38].

Previous studies have primarily focused on the structural properties, electronic band structure, and superconducting transition temperature $T_C$ and thermoelectric (TE) properties of these compounds. Due to the low mass of hydrogen, including zero-point energy (ZPE) contributions can be essential for accurately determining the structural stability of hydrogen-rich compounds [39]. These findings revealed that ternary hydrides are promising candidates in the search for superconductors, potentially synthesizable under mild conditions in hydrogen-rich materials. However, several important physical properties including gravimetric and volumetric hydrogen storage capacities, elastic constants, hardness, anisotropy, sound velocities, thermal and optoelectronic behavior have not yet been investigated. In the present work, we employ the density functional theory (DFT) to explore, for the first time, the structural, electronic, mechanical, hydrogen storage, thermal, and optoelectronic properties of $Mg_2TmH_6$. We have also revisited the superconducting properties for completeness.

## 2. Computational Scheme

Density functional theory (DFT) calculations were performed using the CASTEP code [40]. The plane-wave pseudopotential method [41] was employed in these calculations. The exchange–correlation effects were treated within generalized gradient approximation (GGA) [42]. Ultrasoft pseudopotentials [43] were used to describe the electron–ion interactions. Structural optimization was carried out using the Broyden–Fletcher–Goldfarb–Shanno (BFGS) algorithm [44]. Brillouin zone sampling was performed using the Monkhorst–Pack scheme [45]. The elastic constants were calculated using the stress–strain method as implemented in CASTEP. Thermophysical parameters were subsequently derived from the calculated elastic constants and elastic moduli. The tolerance levels and computational parameters used during the calculations are summarized as follows:

| Parameters | Mg$_2$RhH$_6$ | Mg$_2$PdH$_6$ | Mg$_2$IrH$_6$ | Mg$_2$PtH$_6$ |
|---|---|---|---|---|
| Quality | Ultra-fine | Ultra-fine | Ultra-fine | Ultra-fine |
| Energy (eV/atom) | $5.0 \times 10^{-6}$ | $5.0 \times 10^{-6}$ | $5.0 \times 10^{-6}$ | $5.0 \times 10^{-6}$ |
| Max. force (eV/Å) | 0.01 | 0.01 | 0.01 | 0.01 |
| Max. stress (GPa) | 0.02 | 0.02 | 0.02 | 0.02 |
| Max. displacement (Å) | $5.0 \times 10^{-4}$ | $5.0 \times 10^{-4}$ | $5.0 \times 10^{-4}$ | $5.0 \times 10^{-4}$ |
| Energy cut-off (eV) | 500 | 450 | 600 | 600 |
| $k$-points | 8×8×8 | 10×10×10 | 14×14×14 | 16×16×16 |

## 3. Results and Analysis

### 3.1. Crystal Structure of Mg$_2$TmH$_6$

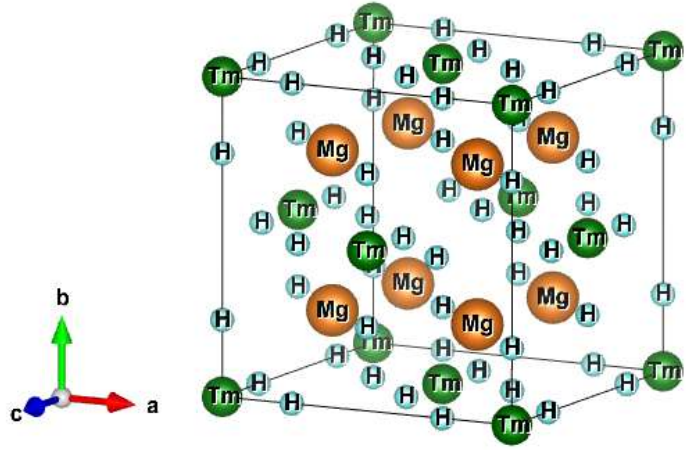

**Fig. 1. Crystal structure of Mg$_2$TmH$_6$ (Tm = Rh, Pd, Ir, Pt).**

The crystal structure of Mg$_2$TmH$_6$ (Tm = Rh, Pd, Ir, Pt) is shown in Fig. 1. In this structure, Mg atoms occupy the 8$c$ Wyckoff site at (0.25, 0.25, 0.25), Tm atoms occupy the 4$a$ site at (0, 0, 0), and H atoms occupy the 24$e$ site at (0.24, 0, 0). Mg$_2$TmH$_6$ crystallizes in the cubic F$m$-3$m$ space group (No. 225). The unit cell contains four formula units, corresponding to a total of 36 atoms per unit cell.

Table 1 summarizes the calculated structural parameters of the cubic Mg$_2$TmH$_6$ (Tm = Rh, Pd, Ir, Pt) hydrides, including the shortest Tm–H bond length, and compared them with previously reported values [34]. The lattice parameter $a$ lies in a narrow range of about 6.63–6.68 Å, increasing slightly from Mg$_2$PdH$_6$ to Mg$_2$PtH$_6$. This trend is reflected in the corresponding cell volumes that vary from roughly 292 Å$^3$ to 298 Å$^3$. The Tm–H bond lengths also show small but systematic differences among the compounds. $Mg_2IrH_6$ exhibits the shortest $Tm$–H distance (≈1.72 Å), while $Mg_2PdH_6$ and $Mg_2PtH_6$ display slightly longer bonds (≈1.79–1.80 Å). Overall, the present results are in very good agreement with the reference data, as indicated by the close correspondence between the values labeled [This] and those from Ref. [34]. This confirms the reliability of the computational approach used for describing the structural properties of these complex hydrides.

**Table 1. Calculated lattice parameter ($a$ in Å) volume ($V$ in Å³), and bond length ($Tm$–H in Å) of $Mg_2TmH_6$ ($Tm$ = Rh, Pd, Ir, Pt).**

| Compounds | a | | V | | ($Tm$-H) | |
|---|---|---|---|---|---|---|
| | [This] | [34] | [This] | [34] | [This] | [34] |
| $Mg_2RhH_6$ | 6.6419 | 6.6540 | 293.01 | 294.61 | 1.727 | 1.700 |
| $Mg_2PdH_6$ | 6.6340 | | 291.96 | | 1.790 | 1.780 |
| $Mg_2IrH_6$ | 6.6665 | | 296.29 | | 1.722 | 1.720 |
| $Mg_2PtH_6$ | 6.6758 | | 297.51 | | 1.801 | 1.790 |

## 3.2. Thermodynamic Stability and Hydrogenation Properties

Since the compounds studied have not been experimentally synthesized, it is essential to assess their thermodynamic stability. To this end, we investigate the cohesive energy per atom ($E_{coh}$) for the first time in this study, which is calculated using the following equation [40–42]:

$$E_{coh} = \frac{E_{Mg_2TmH_6} - 2E_{Mg}^{atom} - E_{Tm}^{atom} - 6E_{H}^{atom}}{9} \qquad (1)$$

where, $E_{Mg_2TmH_6}$ is the total energy per formula unit of $Mg_2TmH_6$. $E_{Mg}$, $E_{Tm}$, and $E_H$ are the total energies of isolated single Mg, $Tm$, and H atoms, respectively. Formation enthalpy can be calculated using the following equation [43–45]:

$$\Delta H = \frac{E_{Mg_2TmH_6} - 2E_{Mg}^{bulk} - E_{Tm}^{bulk} - 6E_{H}^{bulk}}{9} \qquad (2)$$

where, $E_{Mg_2Tm_6}$ is the total energy per formula unit of $Mg_2TmH_6$. $E_{Mg}$, $E_{Tm}$, and $E_H$ are the total energies per atom of bulk Mg, $Tm$, and H, respectively. Our calculated cohesive energy and the formation enthalpy are presented in Table 2. The negative values of $E_{coh}$ and $\Delta H$ indicate the thermodynamic stability of all the compounds.

We have also calculated the gravimetric hydrogen storage capacity and volumetric storage capacity of $Mg_2TmH_6$ ($Tm$ = Rh, Pd, Ir, Pt) for the first time. The gravimetric hydrogen storage capacity ($C_{wt}$) refers to the amount of hydrogen that can be stored per unit mass of the hydrogen storage material. The gravimetric hydrogen storage capacity can be calculated using the following equation [2]:

$$C_{wt} = \left(\frac{n.M_H}{M_{Host} + n.M_H}\right)\% \tag{3}$$

where, $n$ is the mole of hydrogen in formula unit, $M_H$ is the molar mass of hydrogen, and $M_{Host}$ is the molar mass of the non-hydrogen species of the material.

We have calculated volumetric storage capacity using the following equation [2]:

$$\rho_{vol} = \frac{n.M_H}{V(L).N_A} = \frac{40151.83}{Vol.(\text{Å}^3)} \tag{4}$$

where, $n$ is the mole of hydrogen in unit cell, $M_H$ is the molar mass of hydrogen, $V(L)$ is the volume of the unit cell, and $N_A$ is the Avogadro's number. Our estimated values of $C_{wt}$ and $\rho_{vol}$ for all the compounds are presented in Table 2 and graphically are shown in Fig. 2.

$Mg_2IrH_6$ is the most chemically stable compound because it exhibits the lowest cohesive energy (-3.94 eV/atom) and the most exothermic formation enthalpy (-2.55). So, hydrogen is expected to be most strongly bound in this compound. On the other hand, $Mg_2PdH_6$ exhibits the weakest hydrogen binding, suggesting that it may release hydrogen more readily while still maintaining a high volumetric hydrogen density. For gravimetric hydrogen storage capacity, $Mg_2RhH_6$ and $Mg_2PdH_6$ are superior, as they exhibit the highest hydrogen contents by mass (shown in Fig. 2). In contrast, $Mg_2IrH_6$ and $Mg_2IrH_6$ sacrifice gravimetric capacity in favor of stronger metal–hydrogen bonding and slightly lower volumetric hydrogen density. This behavior may be attributed to the higher atomic masses of Ir and Pt, which reduce the hydrogen weight fraction and enhance the strength of metal–hydrogen interactions. Overall, $Mg_2RhH_6$ and $Mg_2PdH_6$ favor higher gravimetric hydrogen capacity, whereas $Mg_2IrH_6$ (and, to a lesser extent, $Mg_2PtH_6$) favors greater thermodynamic stability. The gravimetric storage capacity follows the order: $Mg_2RhH_6$ > $Mg_2PdH_6$ > $Mg_2IrH_6$ > $Mg_2PtH_6$.

**Table 2.** Calculated cohesive energy ($E_{coh}$ in eV/atom), formation enthalpy ($\Delta H$ in eV/atom), gravimetric hydrogen storage capacity ($C_{wt}$ in wt %), and volumetric hydrogen storage capacity ($\rho_{vol}$ in g.H$_2$/L) of Mg$_2$TmH$_6$ (Tm = Rh, Pd, Ir, Pt).

| Compounds | $E_{coh}$ | $\Delta H$ | $C_{wt}$ | $\rho_{vol}$ | Ref. |
|---|---|---|---|---|---|
| Mg$_2$RhH$_6$ | -3.63 | -2.49 | 3.84 | 137.03 | [This] |
| Mg$_2$PdH$_6$ | -3.07 | -2.26 | 3.75 | 137.53 | [This] |
| Mg$_2$IrH$_6$ | -3.94 | -2.55 | 2.45 | 135.52 | [This] |
| Mg$_2$PtH$_6$ | -3.37 | -2.34 | 2.42 | 134.96 | [This] |

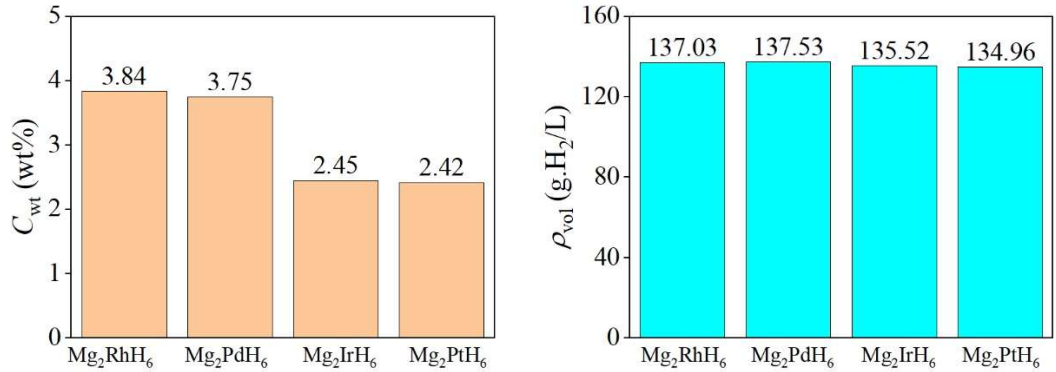

**Fig. 2.** Hydrogen energy storage capacity of Mg$_2$TmH$_6$ (Tm = Rh, Pd, Rr, Pt).

### 3.3. Vibrational properties

The vibrational properties of a material describe the atomic motions about their equilibrium positions and are fundamentally associated with phonons. These properties significantly influence thermal conductivity, heat capacity, phase stability, superconductivity, and optical behavior. The calculated phonon dispersion relations, which represent the vibrational frequencies as a function of wavevector in the Brillouin zone, are presented in Fig. 3 for Mg$_2$TmH$_6$ hydrides.

There is no imaginary phonon frequency present in the entire Brillouin zone, confirming the dynamical stability of these compounds. Furthermore, a clear separation between low-frequency acoustic modes, primarily dominated by heavier Mg and transition-metal atoms, and high-frequency optical modes, mainly associated with hydrogen vibrations, is evident. This distinct distribution reflects the mass difference between constituent atoms and highlights the significant contribution of hydrogen to the high-energy vibrational region.

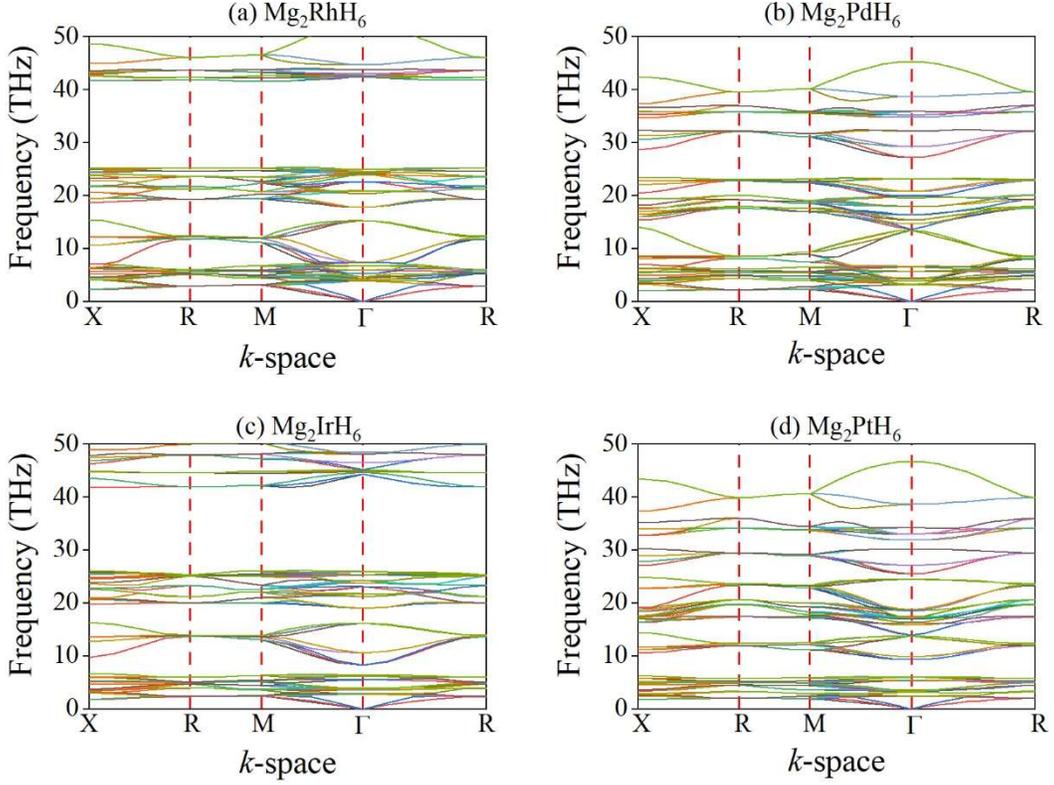

**Fig. 3. Phonon dispersion curve for Mg$_2$$Tm$H$_6$ ($Tm$ = Rh, Pd, Ir, Pt).**

### 3.4. Elastic Properties

#### 3.4.1. Single Crystal Elastic Constants

For cubic crystal, the independent elastic constants are $C_{11}$, $C_{12}$, and $C_{44}$. The calculated elastic constants values are tabulated in Table 3. From Table 3 and Fig. 4(a) it is shown that the elastic constants follow the order $C_{11} > C_{12} > C_{44}$ for all studied hydrides. The constants $C_{11}$, $C_{22}$, and $C_{33}$ represent the resistance to uniaxial stress along [100], [010], and [001] directions, respectively. For cubic crystal, symmetry demands $C_{11} = C_{22} = C_{33}$. This means the incompressibility is same along the three principal axes. Positive values of the elastic constants satisfy the Born stability criteria [46]. For a cubic crystal, the stability conditions are as follows:

$$C_{11} - C_{12} > 0, C_{11} + 2C_{12} > 0, C_{44} > 0 \qquad (5)$$



**Table 3.** Single crystal elastic constants ($C_{ij}$ in GPa) and Cauchy pressure ($P_C$ in GPa) of Mg$_2$TmH$_6$ (Tm = Rh, Pd, Ir, Pt).

| Compounds | $C_{11}$ | $C_{12}$ | $C_{44}$ | $P_C$ | Ref. |
|---|---|---|---|---|---|
| Mg$_2$RhH$_6$ | 120.18 | 50.19 | 23.97 | 26.22 | [This] |
| Mg$_2$PdH$_6$ | 129.61 | 50.27 | 25.15 | 25.12 | [This] |
| Mg$_2$IrH$_6$ | 173.35 | 27.63 | 23.49 | 4.15 | [This] |
| Mg$_2$PtH$_6$ | 139.81 | 57.69 | 31.02 | 26.66 | [This] |

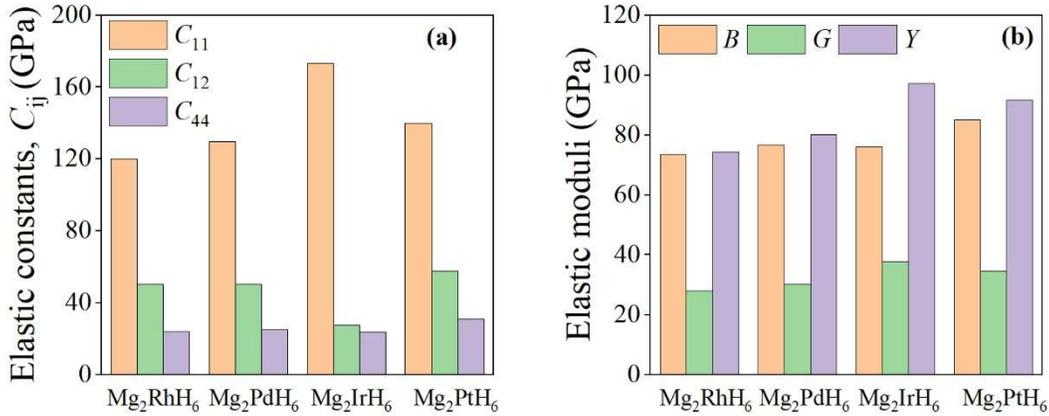

**Fig. 4.** (a) Elastic constants and (b) elastic moduli of Mg$_2$TmH$_6$.

From Table 3, it is observed that $C_{11}$ is greater than $C_{12}$ and $C_{44}$ for all hydrides, indicating that the axial bonding between nearest-neighbor atoms is stronger than the bonding between atoms in different crystal planes. Mg$_2$IrH$_6$ exhibits the strongest axial bonding between nearest-neighbor atoms among all the studied compounds. The Cauchy pressure [$P_C = (C_{12} - C_{44})$] [47] is used to determine the angular character of atomic bindings in solids [48]. Positive Cauchy pressure ($P_C > 0$) usually indicates predominantly metallic bonding and a ductile character, whereas negative value of $P_C$ ($P_C < 0$) is associated with more directional covalent bonding and brittle nature. For Mg$_2$RhH$_6$, Mg$_2$PdH$_6$, Mg$_2$IrH$_6$ and Mg$_2$PtH$_6$, the calculated $P_C$ values are 26.22, 25.12, 4.15, and 26.66 GPa, respectively (cf. Table 3). Since Mg$_2$TmH$_6$ (Tm = Rh, Pd, Ir, Pt) show positive $P_C$, they are expected to possess ductile nature with metallic bonding.

### 3.4.2. Polycrystalline Elastic Properties

Calculated bulk modulus ($B$), shear modulus ($G$), and Young's modulus ($Y$) are displayed in Table 4 as well as in Fig. 4(b). The values of $B$, $G$, and $Y$ are calculated using Voigt [49], Reuss [50], and Hill [51] approximations. Figure 4b shows that, for all Mg$_2$TmH$_6$ compounds studied, the elastic



moduli follow the order Y > B > G. Young's modulus, which represents resistance to length change under applied stress, is the highest modulus for all compounds. $Mg_2PtH_6$ exhibits the largest bulk modulus, indicating the strongest resistance to volume reduction and, consequently, the lowest compressibility among the four hydrides. In contrast, $Mg_2IrH_6$ shows a comparatively higher Young's modulus, suggesting that it is stiffer and more resistant to elastic deformation than $Mg_2RhH_6$, $Mg_2PdH_6$, and $Mg_2PtH_6$.

**Table 4. Calculated polycrystalline elastic moduli (in GPa) of $Mg_2TmH_6$ ($Tm$ = Rh, Pd, Ir, Pt).**

| Compounds | Polycrystalline elastic moduli | | | | | | | | | Ref. |
|---|---|---|---|---|---|---|---|---|---|---|
| | $B_V$ | $B_R$ | $B_H$ | $G_V$ | $G_R$ | $G_H$ | $Y_V$ | $Y_R$ | $Y_H$ | |
| $Mg_2RhH_6$ | 73.52 | 73.52 | 73.52 | 28.38 | 27.43 | 27.91 | 75.44 | 73.19 | 74.32 | [This] |
| $Mg_2PdH_6$ | 76.72 | 76.72 | 76.72 | 30.96 | 29.47 | 30.21 | 81.87 | 78.37 | 80.12 | [This] |
| $Mg_2IrH_6$ | 76.21 | 76.21 | 76.21 | 43.23 | 32.22 | 37.73 | 109.08 | 84.72 | 97.15 | [This] |
| $Mg_2PtH_6$ | 85.06 | 85.06 | 85.06 | 35.04 | 34.39 | 34.71 | 92.43 | 90.91 | 91.67 | [This] |

### 3.4.3. Ductility and Brittleness Indices

The brittle or ductile nature of a solid can be evaluated using Poisson's [52] and Pugh's [53] ratio. The calculated values are listed in Table 5 and shown graphically in Fig. 5. According to Frantsevich [52], a Poisson's ratio ($v$) ≤ 0.26 indicates brittle nature, whereas $v$ > 0.26 indicates ductile behavior. Again, Pugh's ratio ($G/B$) provides a similar classification. If $G/B$ is lower than 0.57 the material is ductile, otherwise it is brittle.

**Table 5. Calculated Pugh's ratio ($G/B$), Poisson's ratio ($v$), and machinability index ($\mu_m$) of $Mg_2TmH_6$ ($Tm$ = Rh, Pd, Ir, Pt).**

| Compounds | $G/B$ | $v$ | $\mu_m$ | Ref. |
|---|---|---|---|---|
| $Mg_2RhH_6$ | 0.38 | 0.33 | 3.07 | [This] |
| $Mg_2PdH_6$ | 0.39 | 0.33 | 3.05 | [This] |
| $Mg_2IrH_6$ | 0.50 | 0.29 | 3.24 | [This] |
| $Mg_2PtH_6$ | 0.41 | 0.32 | 2.74 | [This] |

Fig. 4 shows that all the studied compounds exceed the critical value of Poisson's ratio and fall below the critical value of Pugh's ratio, indicating ductile behavior. Cauchy pressure ($P_C$) is also an indicator of ductile or brittle nature of a material. The values of $P_C$ (listed in Table 3) are consistent with the results obtained from Pugh's ratio and Poisson's ratio. The machinability index $\mu_m$ is defined as $B/C_{44}$. It represents the dry lubricity of the solid. A higher $\mu_m$ means easier



machining and lower frictional loss. Mg$_2$IrH$_6$ shows the highest machinability index and excellent dry lubricity [54]. Overall, the machinability index of the hydrides under investigation are quite high.

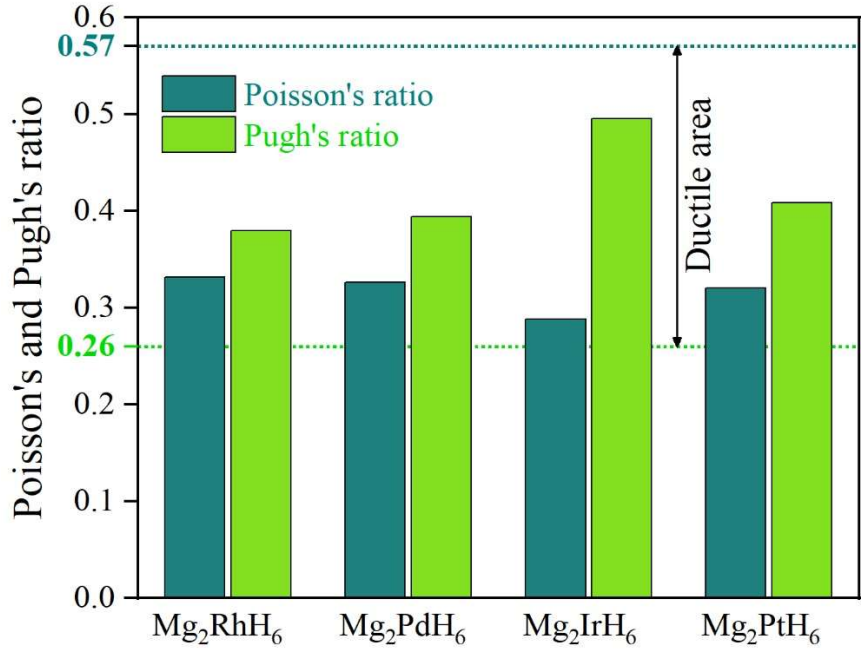

Fig. 5. Poisson's ratio and Pugh's ratio of Mg$_2$TmH$_6$.

### 3.4.4. Elastic Anisotropy Indices

Elastic anisotropy represents the direction-dependent bonding characteristics and mechanical properties of crystalline solids. For an isotropic crystal, the shear elastic anisotropy factor $A$ is equal to unity. In cubic crystals, the single crystal shear anisotropy factors $A$ ($A_1 = A_2 = A_3$) are computed using the following relations [55–57]:

$$A_1 = \frac{4C_{44}}{C_{11} + C_{33} - 2C_{13}} = A_2 = \frac{4C_{55}}{C_{22} + C_{33} - 2C_{23}} = A_3 = \frac{4C_{66}}{C_{11} + C_{22} - 2C_{12}} = A \tag{6}$$

The indices $A^B$ and $A^G$ represent the percentage anisotropies in compressibility and shear, respectively, while $A^U$ denotes the universal anisotropy index. The zero values of $A^B$, $A^G$, and $A^U$ indicate elastic isotropy, whereas non-zero values signify elastic anisotropy [58,59]. The indices $A^B$, $A^G$, and $A^U$ are calculated using the following equations:



$$A^B = \frac{B_V - B_R}{B_V + B_R}, \; A^G = \frac{G_V - G_R}{G_V + G_R}, \text{and } A^U = 5\frac{G_V}{G_R} + \frac{B_V}{B_R} - 6 \geq 0 \qquad (7)$$

The calculated values of $A$, $A^B$, $A^G$ and $A^U$ are listed in Table 6 and graphically represented in Fig. 6. Our estimated values indicate that $Mg_2TmH_6$ are elastically anisotropic. Due to this elastic anisotropy, the crystal binding in $Mg_2TmH_6$ varies in different crystallographic directions.

**Table 6. The shear anisotropy factor ($A = A_1 = A_2 = A_3$), anisotropy indices ($A^B$, $A^G$ in %), and the universal anisotropy indices $A^U$ of $Mg_2TmH_6$ ($Tm$ = Rh, Pd, Ir, Pt).**

| Compounds | $A$ | $A^B$ | $A^G$ | $A^U$ | Ref. |
|---|---|---|---|---|---|
| $Mg_2RhH_6$ | 0.69 | 0.00 | 1.71 | 0.17 | [This] |
| $Mg_2PdH_6$ | 0.63 | 0.00 | 2.47 | 0.25 | [This] |
| $Mg_2IrH_6$ | 0.32 | 0.00 | 14.60 | 1.71 | [This] |
| $Mg_2PtH_6$ | 0.76 | 0.00 | 0.94 | 0.09 | [This] |

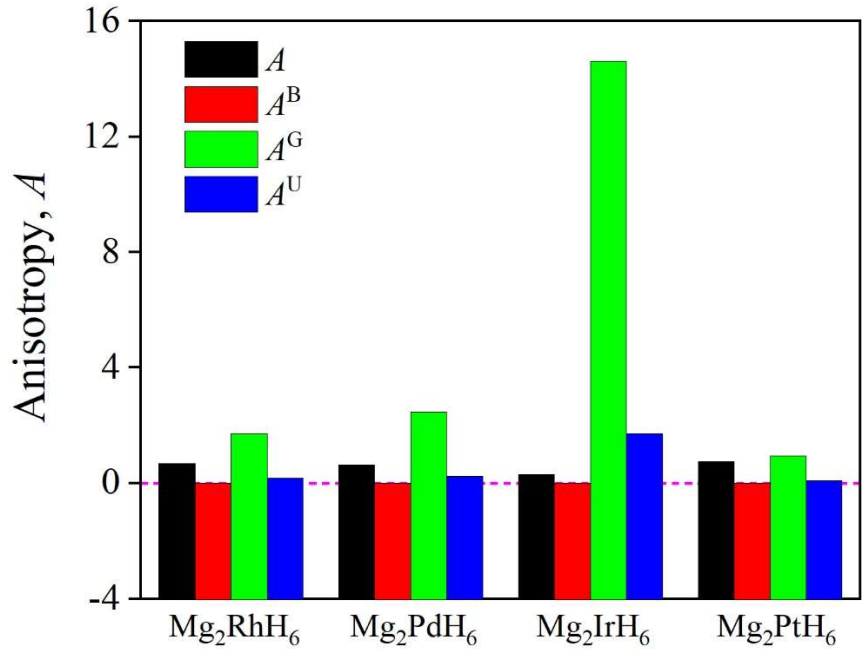

**Fig. 6. Anisotropy indices of $Mg_2TmH_6$.**

Fig. 6 shows that $Mg_2IrH_6$ stands out with significantly higher shear anisotropy ($A^G$) and universal anisotropy ($A^U$), indicating pronounced directional dependence of its elastic properties, especially in shear deformation. The other compounds exhibit relatively low anisotropy indices, suggesting



they are more elastically isotropic. Zero compressibility anisotropy ($A^B$) across all compounds suggests uniform volumetric response to pressure, regardless of crystallographic direction.

### 3.4.5. 3D Visualization of Elastic Anisotropy

The three-dimensional plots of elastic parameters visualize the direction-dependent mechanical behavior of the material. These plots were generated using the ELATE program [60]. Table 7 lists the minimum and maximum values of Young's modulus ($Y$), shear modulus ($G$), linear compressibility ($\beta$), and Poisson's ratio ($v$) for the cubic hydrides $Mg_2RhH_6$, $Mg_2PdH_6$, $Mg_2IrH_6$, and $Mg_2PtH_6$, while Figs. 7–10 represent the corresponding three-dimensional anisotropy surfaces for these quantities.

**Table 7. Minimal and maximal values of Young's modulus ($Y$ in GPa), shear modulus ($G$ in GPa), linear compressibility ($\beta$ in TPa$^{-1}$), and Poisson's ratio ($v$) for $Mg_2TmH_6$ ($Tm$ = Rh, Pd, Ir, Pt).**

| Compounds | $Y_{min}$ | $Y_{max}$ | $G_{min}$ | $G_{max}$ | $\beta_{min}$ | $\beta_{max}$ | $v_{min}$ | $v_{max}$ | Ref. |
|---|---|---|---|---|---|---|---|---|---|
| $Mg_2RhH_6$ | 64.87 | 90.61 | 23.97 | 34.99 | 4.534 | 4.534 | 0.227 | 0.456 | [This] |
| $Mg_2PdH_6$ | 68.03 | 101.51 | 25.15 | 39.67 | 4.345 | 4.345 | 0.204 | 0.474 | [This] |
| $Mg_2IrH_6$ | 63.89 | 165.75 | 23.49 | 72.86 | 4.374 | 4.374 | 0.063 | 0.607 | [This] |
| $Mg_2PtH_6$ | 82.98 | 106.12 | 31.02 | 41.06 | 3.919 | 3.919 | 0.242 | 0.415 | [This] |

From Fig. 7 the three-dimensional representation of Young's modulus is nearly spherical for $Mg_2RhH_6$ and $Mg_2PtH_6$, indicating low anisotropy and almost uniform stiffness in all directions. For $Mg_2PdH_6$, the surface becomes more elongated, revealing a preferred stiff direction. In contrast, $Mg_2IrH_6$ exhibits a strongly lobed, star-like shape, confirming the largest stiffness contrast among different crystallographic directions.

The 3D surface of linear compressibility (Fig. 8) exhibits an almost perfect spherical shape, indicating negligible elastic anisotropy and nearly uniform stiffness along all crystallographic directions for all the hydrides studied. This behavior is further confirmed by the identical values of $\beta_{min}$ and $\beta_{max}$ for all compounds, as listed in Table 7.

Fig. 9 shows that the shear modulus surface of $Mg_2RhH_6$ is relatively rounded with moderately developed lobes, while $Mg_2PtH_6$ exhibits only weak lobing, both indicating mild to moderate shear anisotropy. In contrast, $Mg_2PdH_6$ displays a strongly distorted, multi-lobed three-dimensional shear modulus surface, reflecting pronounced elastic anisotropy and a strong directional



dependence of shear stiffness. The most highly petaled and complex multi-lobed surface is observed for Mg$_2$IrH$_6$, indicating the highest degree of shear anisotropy among the studied compounds.

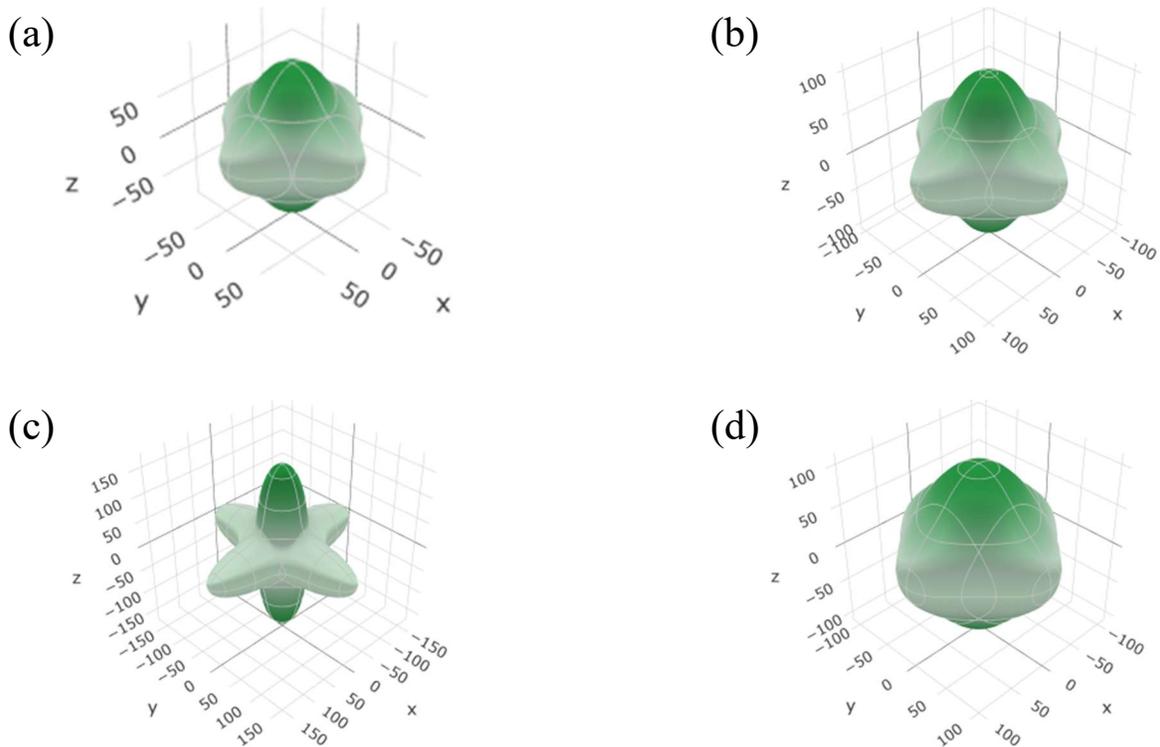

Fig. 7. 3D view of Young's modulus for (a) Mg$_2$RhH$_6$, (b) Mg$_2$PdH$_6$, (c) Mg$_2$IrH$_6$, and (d) Mg$_2$PtH$_6$.

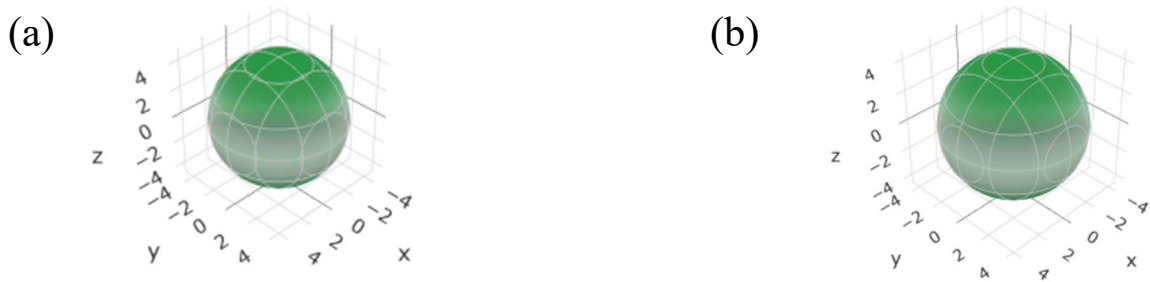



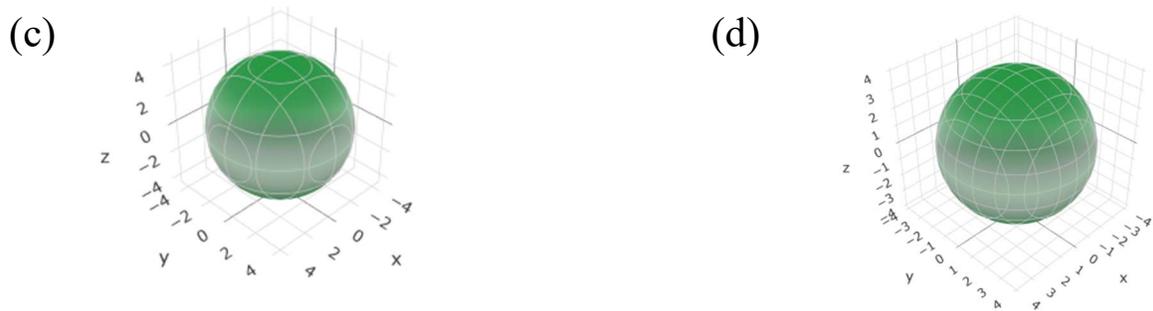

**Fig. 8.** 3D view of linear compressibility for (a) $Mg_2RhH_6$, (b) $Mg_2PdH_6$, (c) $Mg_2IrH_6$, and (d) $Mg_2PtH_6$.

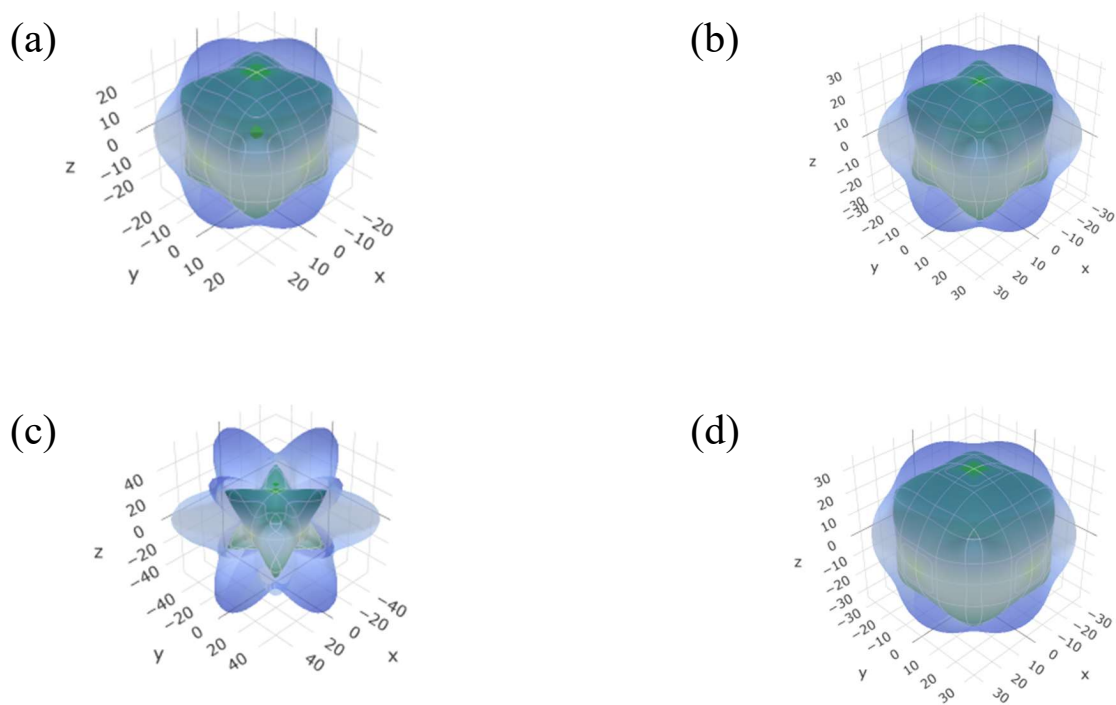

**Fig. 9.** 3D view of shear modulus for (a) $Mg_2RhH_6$, (b) $Mg_2PdH_6$, (c) $Mg_2IrH_6$, and (d) $Mg_2PtH_6$.

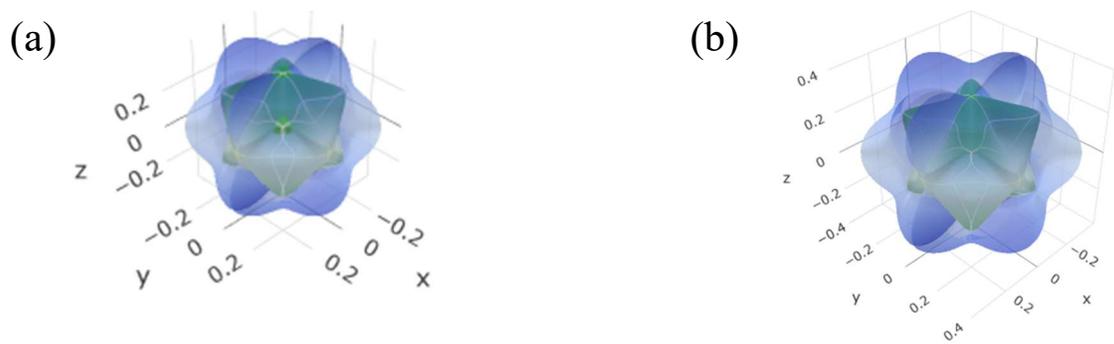



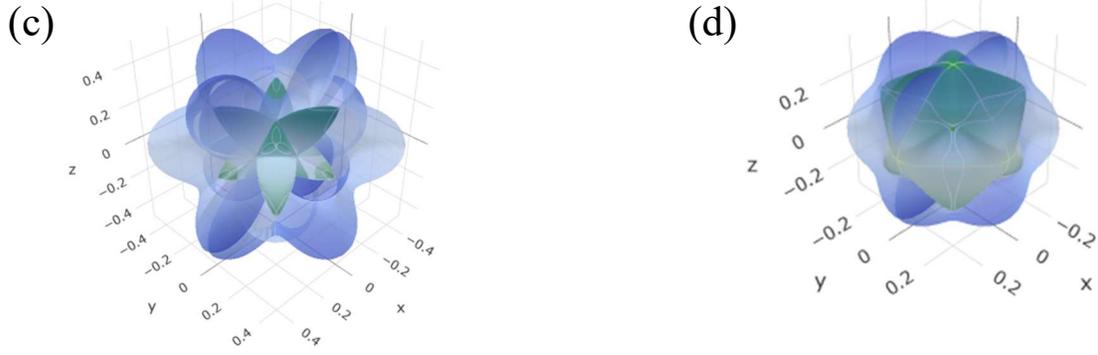

Fig. 10. 3D view of Poisson's ratio for (a) Mg$_2$RhH$_6$, (b) Mg$_2$PdH$_6$,

(c) Mg$_2$IrH$_6$, and (d) Mg$_2$PtH$_6$.

Fig. 10 shows that the Poisson's ratio stays smooth and nearly spherical for Mg$_2$RhH$_6$ and Mg$_2$PtH$_6$, indicating a stable and weakly direction-dependent transverse strain response. Mg$_2$PdH$_6$ exhibits a mildly petaled but still smooth surface, consistent with modest angular variation. In contrast, Mg$_2$IrH$_6$ displays a strongly lobed surface, reflecting large directional changes in the lateral-to-axial strain ratio. Overall, the degree of anisotropy increases from Mg$_2$PtH$_6$ and Mg$_2$RhH$_6$ to Mg$_2$PdH$_6$, reaching its maximum in Mg$_2$IrH$_6$.

### 3.5. Acoustic Properties and Material Hardness

#### 3.5.1. Sound Velocities

The sound velocities in the hydride materials were calculated using the following equations [66–69] and are listed in Table 8.

$$v_t = \sqrt{\frac{G}{\rho}}, \quad v_l = \sqrt{\frac{3B + 4G}{3\rho}}, \text{ and } v_m = \left[\frac{1}{3}\left(\frac{2}{v_t^3} + \frac{1}{v_l^3}\right)\right]^{-\frac{1}{3}} \tag{8}$$

where, $v_t, v_l$, and $v_m$ signify the transverse, longitudinal, and the mean sound velocities, respectively. There are many properties such as crystal stiffness, crystal density, Debye temperature and thermophysical properties, which are closely related with the sound velocities [61]. Crystalline solids support both longitudinal and transverse modes of propagation of acoustic disturbances.



**Table 8.** Calculated density ($\rho$ in gm/cm$^3$), transverse sound velocities ($v_t$ in km/s), longitudinal sound velocities ($v_l$ in km/s), and average sound velocities ($v_m$ in km/s) of Mg$_2$TmH$_6$ (Tm = Rh, Pd, Ir, Pt).

| Compounds | $\rho$ | $v_t$ | $v_l$ | $v_m$ | Ref. |
|---|---|---|---|---|---|
| Mg$_2$RhH$_6$ | 3.572 | 2.8 | 5.6 | 3.1 | [This] |
| Mg$_2$PdH$_6$ | 3.664 | 2.9 | 5.7 | 3.2 | [This] |
| Mg$_2$IrH$_6$ | 5.535 | 2.69 | 4.8 | 2.9 | [This] |
| Mg$_2$PtH$_6$ | 5.576 | 2.5 | 4.9 | 2.8 | [This] |

Table 8 summarizes the calculated density ($\rho$) and acoustic sound velocities for the Mg$_2$TmH$_6$ (Tm = Rh, Pd, Ir, Pt). The densities span from 3.572–3.664 gm/cm$^3$ for Mg$_2$RhH$_6$ and Mg$_2$PdH$_6$ to markedly higher values of 5.535–5.576 gm/cm$^3$ for Mg$_2$IrH$_6$ and Mg$_2$PtH$_6$, reflecting the larger atomic masses of Ir and Pt. Fig. 11 shows that the transverse sound velocities are highest for Mg$_2$PdH$_6$ (2.9 km/s) and Mg$_2$RhH$_6$ (2.8 km/s), but decrease for the heavier compounds Mg$_2$IrH$_6$ (2.6 km/s) and Mg$_2$PtH$_6$ (2.5 km/s). A similar trend is observed for longitudinal velocities, which fall from ~5.6 km/s in the Mg$_2$RhH$_6$ and Mg$_2$PdH$_6$ compounds to ~4.8 km/s in the Mg$_2$IrH$_6$ and Mg$_2$PtH$_6$ compounds. Consequently, the average sound velocity follows the order $v_m$(Mg$_2$PdH$_6$) > $v_m$(Mg$_2$RhH$_6$) > $v_m$(Mg$_2$IrH$_6$) > $v_m$(Mg$_2$PtH$_6$). These systematic reductions in $v_t, v_l,$ and $v_m$ with increasing density indicate softer elastic response and lower lattice vibrational speeds for Mg$_2$IrH$_6$ and Mg$_2$PtH$_6$ hydrides.

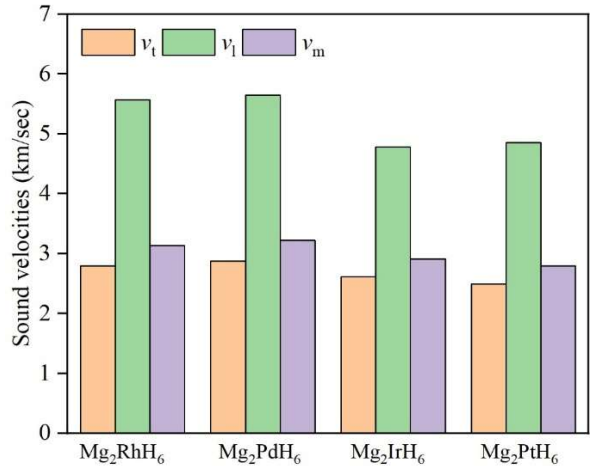

**Fig. 11.** Sound velocities for Mg$_2$TmH$_6$ (Tm = Rh, Pd, Ir, Pt).

The ultrasonic sound velocities can be categorized into three types one longitudinal $v_l$ and two shear $v_{t_1}$ and $v_{t_2}$. We have also calculated these three types of ultrasonic sound velocities using the following equations [62].



For [100]:

$$v_{l\,[100]} = \sqrt{\frac{C_{11}}{\rho}};\ v_{t_1[100]} = v_{t_2[100]} = \sqrt{\frac{C_{44}}{\rho}} \tag{9}$$

For [110]:

$$v_{l\,[110]} = \sqrt{\frac{C_{11} + C_{12} + 2C_{44}}{2\rho}};\ v_{t_1[110]} = \sqrt{\frac{C_{44}}{\rho}};\ v_{t_2[100]} = \sqrt{\frac{C_{11} - C_{12}}{2\rho}} \tag{10}$$

For [111]:

$$v_{l\,[111]} = \sqrt{\frac{C_{11} + 2C_{12} + 4C_{44}}{3\rho}};\ v_{t_1[111]} = v_{t_2[111]} = \sqrt{\frac{C_{11} - C_{12} + C_{44}}{3\rho}} \tag{11}$$

The direction dependent sound velocities are listed in Table 9. The longitudinal sound velocities are greater than the transverse velocities in all propagation directions, due to $C_{11}$ being larger than $C_{12}$ or $C_{44}$ (Table 3) for Mg$_2$TmH$_6$. The direction-dependent sound velocities of the studied compounds also reflect their elastic anisotropy.

**Table 9.** The longitudinal ($v_l$ in m/s) and transverse wave velocities ($v_t, v_{t_1}$, and $v_{t_2}$ in m/s) along [100], [110], and [111] directions for Mg$_2$TmH$_6$ (Tm = Rh, Pd, Ir, Pt).

| Compounds | [100] | | [110] | | | [111] | | Ref. |
|---|---|---|---|---|---|---|---|---|
| | $v_l$ | $v_t$ | $v_l$ | $v_{t_1}$ | $v_{t_2}$ | $v_l$ | $v_t$ | |
| Mg$_2$RhH$_6$ | 5.8 | 2.6 | 5.5 | 2.6 | 3.1 | 5.4 | 3.0 | [This] |
| Mg$_2$PdH$_6$ | 5.9 | 2.6 | 5.6 | 2.6 | 3.3 | 5.5 | 3.1 | [This] |
| Mg$_2$IrH$_6$ | 5.6 | 2.1 | 4.7 | 2.1 | 3.6 | 4.4 | 3.2 | [This] |
| Mg$_2$PtH$_6$ | 5.0 | 2.4 | 4.8 | 2.4 | 2.7 | 4.8 | 2.6 | [This] |

### 3.5.2. Hardness

Hardness quantifies elastic and plastic behavior of a solid. Here we have calculated the hardness of Mg$_2$TmH$_6$ (Tm = Rh, Pd, Ir, Pt) using the different formalisms developed by Teter *et al*. [63], Tian *et al*. [64], Chen *et al*. [65], the microhardness [66], and that by Efim Mazhnik [67]. The calculated values of hardness are listed in Table 10 and illustrated in Fig. 12.



**Table 10. Calculated hardness ($H$ in GPa) of Mg$_2$TmH$_6$ (Tm = Rh, Pd, Ir, Pt).**

| Compounds | Hardness $H$ | | | | | Ref. |
|---|---|---|---|---|---|---|
| | $H_{Teter}$ | $H_{Tian}$ | $H_{Chen}$ | $H_{micro}$ | $H_{Efim\ Mazhnik}$ | |
| Mg$_2$RhH$_6$ | 4.21 | 2.58 | 1.51 | 3.13 | 4.11 | [This] |
| Mg$_2$PdH$_6$ | 4.56 | 2.87 | 1.94 | 3.51 | 4.40 | [This] |
| Mg$_2$IrH$_6$ | 5.70 | 4.59 | 4.35 | 5.34 | 4.97 | [This] |
| Mg$_2$PtH$_6$ | 5.24 | 3.32 | 2.58 | 4.16 | 4.99 | [This] |

Considering all formalisms, Mg$_2$IrH$_6$ demonstrates the highest mechanical hardness, reflecting its superior resistance to plastic deformation compared to the other compounds (Fig. 12). Mg$_2$PtH$_6$ possesses slightly lower but still substantial hardness, whereas Mg$_2$PdH$_6$ and especially Mg$_2$RhH$_6$ are the softest members of this hydride family, consistent across all formalisms. Fig. 12 further visualizes these trends, with the bars for Mg$_2$IrH$_6$ and Mg$_2$PtH$_6$ systematically taller than those for Mg$_2$PdH$_6$ and Mg$_2$RhH$_6$, demonstrating that an appropriate choice of transition metal can effectively tune the mechanical robustness of Mg$_2$TmH$_6$ hydrides. Despite these variations, the relatively small hardness values indicate that all Mg$_2$TmH$_6$ compounds can be easily compressed, cut, bent, or scratched.

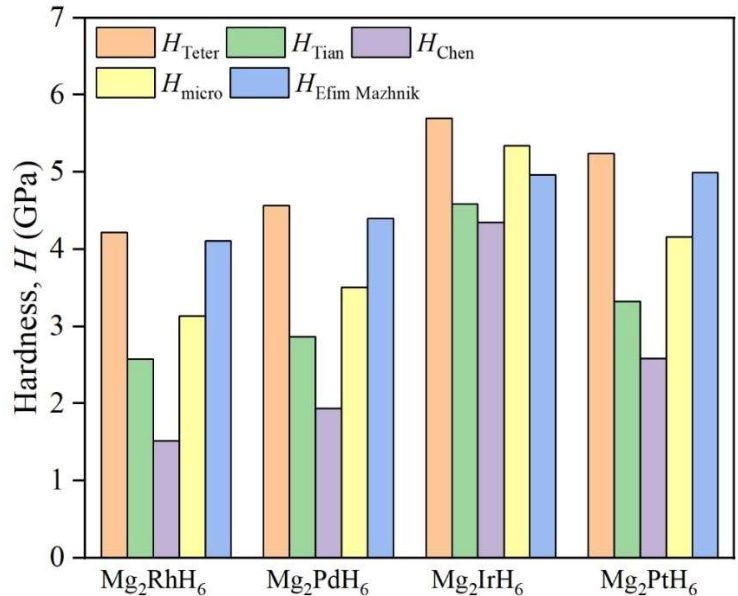

**Fig. 12. Hardness of Mg$_2$TmH$_6$ (Tm = Rh, Pd, Ir, Pt).**

### 3.6. Thermophysical Properties

Few important thermophysical properties of Mg$_2$TmH$_6$ (Tm = Rh, Pd, Ir, Pt) are investigated, and the calculated values are listed in Table 11.



Among these, the Debye temperature $\theta_D$ is closely related to the specific heat, melting temperature, thermal conductivity, hardness of solids, elastic constants, acoustic velocity, resistivity etc. The value of Debye temperature gives information about the superconducting transition temperature in electron-phonon superconductors. At low temperatures, lattice vibrational excitations arise solely from the acoustic modes. Thus, the Debye temperature calculated by elastic constants is considered to be close to that acquired from the specific heat measurements.

The expression of Debye temperature from the Anderson method [68] is as follows:

$$\theta_D = \frac{h}{k_B} \left[ \left( \frac{3n}{4\pi} \right) \frac{N_A \rho}{M} \right]^{1/3} v_m \qquad (12)$$

where, $h$ is Planck's constant, $k_B$ is the Boltzmann's constant, $\rho$ is the density, $N_A$ is the Avogadro number, $M$ is the molecular mass, and $v_m$ is the average sound velocity. The Debye temperatures of cubic $Mg_2TmH_6$ hydrides show moderate variation, reflecting differences in lattice stiffness and characteristic phonon frequencies. $Mg_2PdH_6$ exhibits the highest $\theta_D$ value of 476.92 K, closely followed by $Mg_2RhH_6$ with 464.02 K, indicating relatively strong interatomic bonding and higher sound velocities in these compounds. In contrast, $Mg_2IrH_6$ and $Mg_2PtH_6$ display somewhat lower Debye temperatures of 429.39 K and 411.50 K, respectively, suggesting slightly softer lattices and lower average vibrational frequencies. Overall, the trend $\theta_D$ ($Mg_2PdH_6$) > $\theta_D$ ($Mg_2RhH_6$) > $\theta_D$ ($Mg_2IrH_6$) > $\theta_D$ ($Mg_2PtH_6$) implies that $Mg_2PdH_6$ and $Mg_2RhH_6$ hydrides possess the stiffest phonon spectra among the series.

Table 11. Calculated Debye temperature ($\theta_D$ in K), thermal expansion coefficient ($\alpha$ in $10^{-5}$ K$^{-1}$), melting temperature ($T_m$ in K), minimum thermal conductivity ($k_{min}$ in W m$^{-1}$ K$^{-1}$), Grüneisen parameter ($\gamma$), and Kleinman parameter ($\zeta$) of cubic $Mg_2TmH_6$ ($Tm$ = Rh, Pd, Ir, Pt).

| Compounds | $\theta_D$ | $\alpha$ | $T_m$ | $\gamma$ | $k_{min}$ Cahill | $k_{min}$ Clarke | $\zeta$ | Ref. |
|---|---|---|---|---|---|---|---|---|
| $Mg_2RhH_6$ | 464.02 | 5.73 | 1263.28 | 1.99 | 1.54 | 1.07 | 0.55 | [This] |
| $Mg_2PdH_6$ | 476.92 | 5.30 | 1319.01 | 1.95 | 1.57 | 1.10 | 0.53 | [This] |
| $Mg_2IrH_6$ | 429.39 | 4.24 | 1577.51 | 1.70 | 1.37 | 0.99 | 0.31 | [This] |
| $Mg_2PtH_6$ | 411.50 | 4.61 | 1379.31 | 1.91 | 1.34 | 0.94 | 0.55 | [This] |

The thermal expansion coefficients (TEC) indicate how sensitively the lattice parameters of a compound respond to temperature. The TEC is denoted by $\alpha$. In this work the TEC of the material was calculated using the following equation [69]:



$$\alpha = \frac{1.6 \times 10^{-3}}{G} \tag{13}$$

The relation between thermal expansion coefficient and melting temperature is approximated as $\alpha \approx 0.02/T_m$ [70], which indicates that the thermal expansion coefficient decreases as the melting temperature increases. This implies that materials with a higher melting point tend to exhibit a smaller thermal expansion. Our computed results follow this expected inverse trend and therefore align well with the stated relationship. From Table 11, it is shown that $Mg_2RhH_6$ has the largest $\alpha$ (5.73×10$^{-5}$ K$^{-1}$), followed by $Mg_2PdH_6$ (5.30×10$^{-5}$ K$^{-1}$) implying stronger lattice dilatation with heating for these two compounds. On the other hand, $Mg_2PtH_6$ and $Mg_2IrH_6$ exhibit lower expansion coefficients of 4.61×10$^{-5}$ K$^{-1}$ and 4.24×10$^{-5}$ K$^{-1}$, respectively, meaning they are more dimensionally stable under thermal loading. The overall decrease of $\alpha$ from $Mg_2RhH_6$ and $Mg_2PdH_6$ to $Mg_2IrH_6$ and $Mg_2PtH_6$ suggests that the latter hydrides maintain structural integrity better at elevated temperatures.

The melting temperature is mostly used to gauge the overall bonding strength and limit of high temperature applicability of solids. The melting temperature of cubic compound is calculated with an empirical formula based on the single crystal elastic constants as follows [70]:

$$T_m = 553 + 5.91 C_{11} \tag{14}$$

The calculated $T_m$ is listed in Table 11. In general, compounds with high melting temperature have lower thermal expansion coefficient. The estimated melting temperatures further highlight the thermal stability of $Mg_2TmH_6$ hydrides. From Table 11, it is observed that $Mg_2IrH_6$ possesses the highest melting temperature of 1577.51 K, indicating the greatest resistance to thermal breakdown and making it the most thermally robust compound in the series. $Mg_2PtH_6$ also shows a relatively high melting point of 1379.31 K, while $Mg_2PdH_6$ and $Mg_2RhH_6$ have somewhat lower melting values of 1319.01 K and 1263.28 K, respectively. This trend indicates that substituting Pd or Rh with Ir or Pt improves the high-temperature stability of $Mg_2TmH_6$ hydrides, making them more suitable for applications that require operation at elevated temperatures.

The Grüneisen parameter provides insight into lattice anharmonicity and the strength of phonon-phonon interactions. The Grüneisen parameter is estimated as follows: $\gamma = \frac{3}{2} \frac{1+\nu}{2-3\nu}$ [71]. The values



of Poisson's ratio i.e., $v = -1.0$ corresponds to a completely harmonic solid. High value of Grüneisen parameter is sometimes indicative of strong electron-phonon interaction in phonon mediated superconductors [72]. The obtained values of $\gamma$ are typical [73]. From Table 11, it is observed that among the Mg$_2$TmH$_6$ hydrides, Mg$_2$RhH$_6$ exhibits the largest $\gamma$ value (1.99), followed closely by Mg$_2$PdH$_6$ (1.95) and Mg$_2$PtH$_6$ (1.91), indicating relatively stronger anharmonic effects and larger sensitivity of vibrational frequencies to volume changes. In contrast, Mg$_2$IrH$_6$ exhibits the lowest $\gamma$ value (1.70), suggesting comparatively weaker anharmonicity and more harmonic lattice vibrations. These variations imply that thermal expansion and thermal conductivity are more strongly temperature dependent in Mg$_2$RhH$_6$, Mg$_2$PdH$_6$, and Mg$_2$PtH$_6$ hydrides than in Mg$_2$IrH$_6$ compound.

At temperatures above the Debye temperature, thermal conductivity of solids attains a minimum saturating value, known as the minimum thermal conductivity ($k_{min}$). According to the Cahill model, minimum thermal conductivity can be calculated using the following equation [74]:

$$k_{min} = \frac{k_B}{2.48} n^{\frac{2}{3}} (v_l + v_{t1} + v_{t2}) \tag{15}$$

From this expression it is clear that minimum thermal conductivity depends on sound velocities in different propagation modes. Clarke also suggested the following equation for $k_{min}$ [75]:

$$k_{min} = k_B v_m (V_{atomic})^{-\frac{2}{3}} \tag{16}$$

where, $k_B$ is the Boltzmann constant, $v_m$ is the average sound velocities, and $V_{atomic}$ is the cell volume per atom.

The minimum thermal conductivities estimated using the Cahill and Clarke models exhibit only modest variation across the Mg$_2$TmH$_6$ family. Within the Cahill framework, $k_{min}$ ranges from 1.34 W m$^{-1}$ K$^{-1}$ for Mg$_2$PtH$_6$ and 1.37 W m$^{-1}$ K$^{-1}$ for Mg$_2$IrH$_6$ to slightly higher values of 1.54 and 1.57 W m$^{-1}$ K$^{-1}$ for Mg$_2$RhH$_6$ and Mg$_2$PdH$_6$, respectively. The Clarke model yields consistently lower values while preserving the same ordering, with $k_{min}$ ranging from 0.94–0.99 W m$^{-1}$ K$^{-1}$ for Mg$_2$PtH$_6$ and Mg$_2$IrH$_6$ to 1.07–1.10 W m$^{-1}$ K$^{-1}$ for Mg$_2$RhH$_6$ and Mg$_2$PdH$_6$. These results indicate that Mg$_2$PdH$_6$ and Mg$_2$RhH$_6$ can sustain relatively more efficient phonon-mediated heat transport, whereas Mg$_2$IrH$_6$ and Mg$_2$PtH$_6$ act as slightly poorer thermal conductors, a characteristic that may be advantageous for thermal insulation in certain device applications.



The Kleinman parameter ($\zeta$) quantifies the nature of internal strain and was introduced by Kleinman [76] to describe the relative ease of bond bending versus the bond stretching. Low level of bond bending contribution leads to $\zeta = 0$; while low level of bond stretching leads to $\zeta = 1$. The Kleinman parameter calculated from the elastic constants $C_{11}$ and $C_{12}$ is as follows:

$$\zeta = \frac{C_{11} + 8C_{12}}{7C_{11} + 2C_{12}} \tag{17}$$

The values of $\zeta$ of Mg$_2$*Tm*H$_6$ are given in Table 11. For Mg$_2$RhH$_6$, Mg$_2$PdH$_6$, and Mg$_2$PtH$_6$, the parameter $\zeta$ falls within a narrow range of 0.53–0.55, indicating a similar balance between bond-stretching and bond-bending contributions to their deformation behavior. In contrast, Mg$_2$IrH$_6$ exhibits a markedly lower $\zeta$ value of 0.31, implying a distinct internal strain response in which bond stretching is comparatively more favorable than bond bending. This pronounced difference suggests that the bonding geometry and mechanical response of Mg$_2$IrH$_6$ under applied stress differ significantly from those of the other transition-metal hydrides considered in this study.

### 3.7. Electronic Properties

#### 3.7.1. Density of States

Electronic energy density of states (DOS), derived from the electronic band structure (Section 3.7.2) represents the number of electronic states available per unit energy at each energy level, providing key insights into its electronic, optical, and thermal properties. Partial density of states (PDOS) is the projection of the total density of states onto specific atoms or orbitals, showing their contribution to the electronic states. It helps reveal bonding characteristics, orbital hybridization, and the role of individual elements in the material's electronic structure.

The total and partial densities of states (TDOS and PDOS, respectively) are shown in Fig. 13. The values of TDOS at the Fermi level are 1.93, 1.43, 1.54 and 1.19 states/eV for Mg$_2$RhH$_6$, Mg$_2$PdH$_6$, Mg$_2$IrH$_6$ and Mg$_2$PtH$_6$, respectively. The non-zero DOS at Fermi level confirms the metallic nature of Mg$_2$*Tm*H$_6$. At the Fermi level, these TDOS values mainly originated from H-*s*, *Tm*-*d* and Mg-*p* states. Hybridization among the orbitals of the constituent atoms (Mg, *Tm*, H) suggests covalent bonding in Mg$_2$*Tm*H$_6$. The pseudogap, located close to the Fermi level, suggests high electronic



stability of the compounds studied. The TDOS at the Fermi level, $N(E_F)$, is used to estimate the superconducting properties of $Mg_2TmH_6$ metals in Section 3.8.

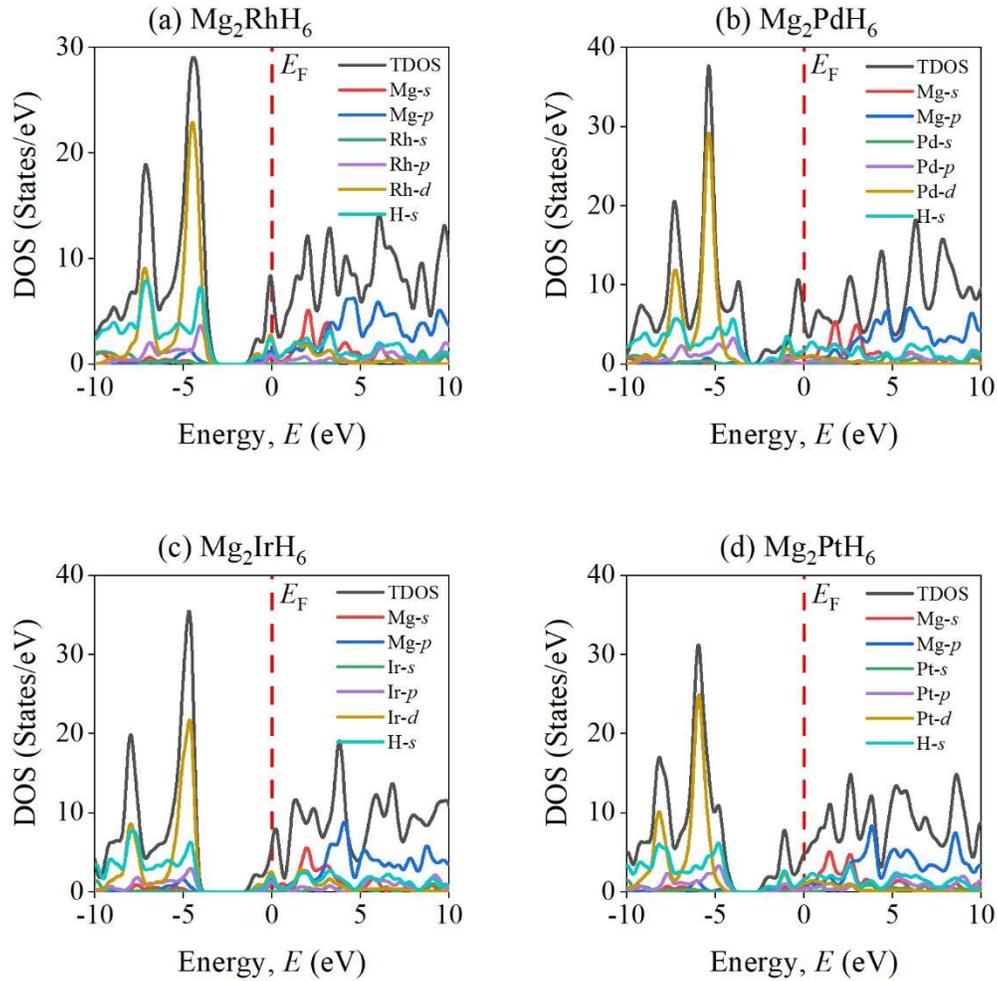

**Fig. 13. Electron density of states of $Mg_2TmH_6$ ($Tm$ = Rh, Pd, Ir, Pt).**

### 3.7.2. Electronic Band Structure

The calculated electronic band structures of $Mg_2TmH_6$ are illustrated in Fig. 14. As seen in Fig. 14, no band gap is present, indicating the metallic nature of all $Mg_2TmH_6$ compounds. The number of bands crossing the Fermi level is two for $Mg_2RhH_6$ and $Mg_2IrH_6$, five for $Mg_2PdH_6$, and nine for $Mg_2PtH_6$. Consequently, $Mg_2PtH_6$ is expected to exhibit the highest electrical conductivity among the hydrides investigated in this work. The most prominent bands crossing the Fermi level,



for all these hydrides running along M-Γ-R are highly dispersive. Thus, the charge effective mass is expected to be low resulting in high mobility.

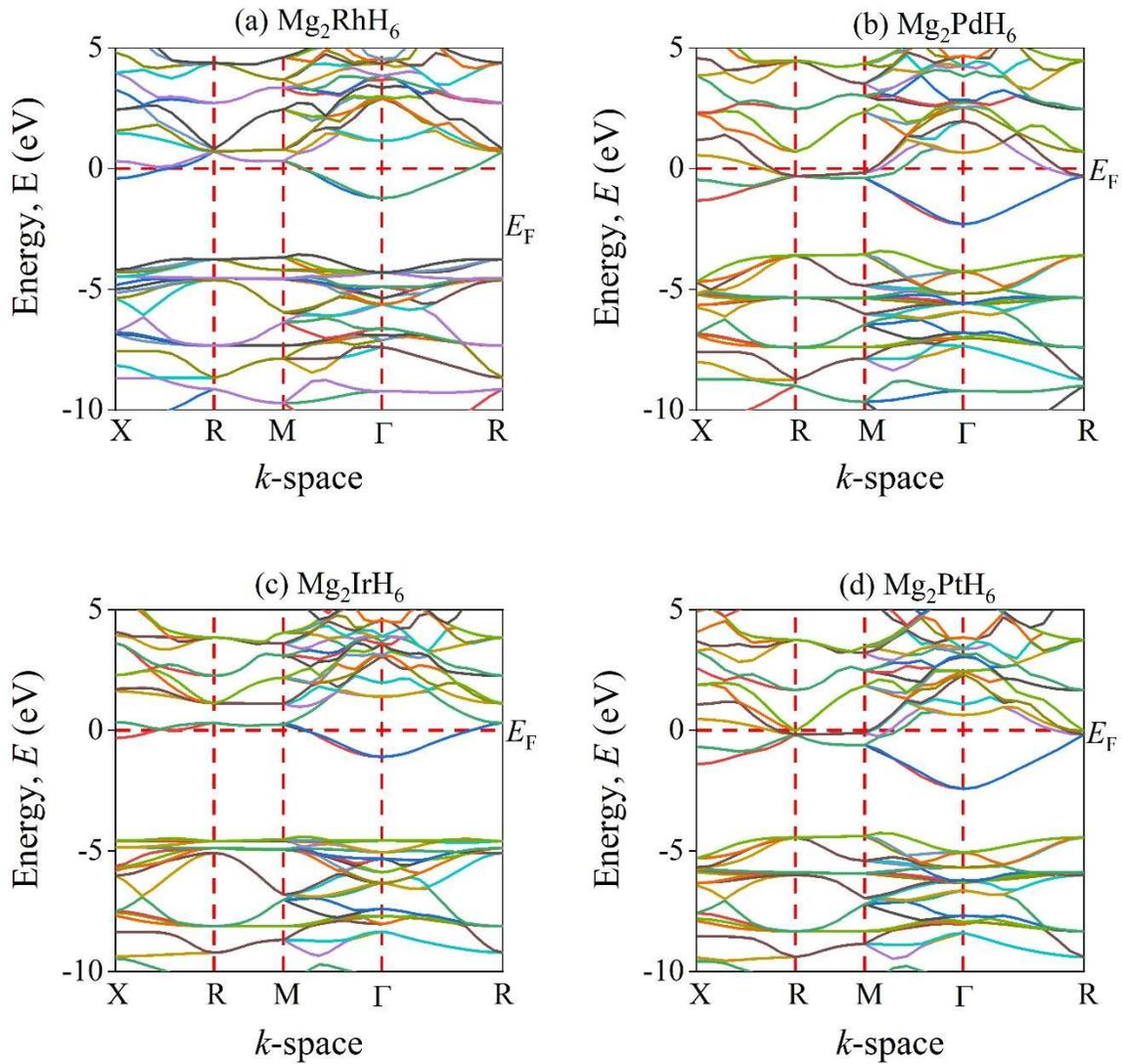

**Fig. 14. Electronic band structure of Mg$_2$*Tm*H$_6$ (*Tm* = Rh, Pd, Ir, Pt).**

### 3.7.3. Mulliken Atomic Populations

Mulliken atomic populations and Hirshfeld charges for the series of complex hydrides Mg$_2$*Tm*H$_6$ (*Tm* = Rh, Pd, Ir, Pt) are listed in Tables 12 and 13, respectively. Each compound consists of Mg, a transition metal (Tm), and H atoms. The tables illustrate how the electronic charge is distributed among these constituent atoms within the crystal structure.



Table 12 summarizes the Mulliken population analysis for the Mg$_2$TmH$_6$ (Tm = Rh, Pd, Ir, Pt) compounds, including the charge spilling parameter and the orbital-resolved electron populations for each atomic species. For each compound, the table reports the number of ions considered, the occupation of the *s*, *p*, *d* and *f* orbitals, the total electron population, and the resulting Mulliken atomic charges (difference between the neutral atom valence electron count and the total Mulliken population). Positive Mulliken charges indicate that an atom has lost electronic charge and behaves as a cation, whereas negative charges indicate that an atom has gained electronic charge and behaves as an anion. The charge spilling parameter (%) reported for each compound quantifies the fraction of electronic charge not properly described within the chosen basis set. The very small values observed (0.25–0.40%) demonstrate that the basis set is adequate and that the population analysis is reliable for qualitative discussion of charge distribution.

**Table 12. Mulliken atomic populations for Mg$_2$TmH$_6$.**

| Compounds | Charge spilling parameter (%) | Mulliken Populations | | | | | | | | Ref. |
|---|---|---|---|---|---|---|---|---|---|---|
| | | Species | No. of Ions | s | p | d | f | Total | Charge (e) | |
| Mg$_2$RhH$_6$ | 0.36 | Mg | 8 | 0.48 | 5.88 | 0.00 | 0.00 | 6.36 | 1.64 | [This] |
| | | Rh | 4 | 0.96 | 1.54 | 8.40 | 0.00 | 10.91 | -1.91 | |
| | | H | 24 | 1.23 | 0.00 | 0.00 | 0.00 | 1.23 | -0.23 | |
| Mg$_2$PdH$_6$ | 0.25 | Mg | 8 | 0.30 | 5.87 | 0.00 | 0.00 | 6.17 | 1.83 | [This] |
| | | Pd | 4 | 1.03 | 2.10 | 9.36 | 0.00 | 12.48 | -2.48 | |
| | | H | 24 | 1.20 | 0.00 | 0.00 | 0.00 | 1.20 | -0.20 | |
| Mg$_2$IrH$_6$ | 0.40 | Mg | 8 | 0.44 | 6.04 | 0.00 | 0.00 | 6.48 | 1.52 | [This] |
| | | Ir | 4 | 0.87 | 1.73 | 8.11 | 0.00 | 10.70 | -1.70 | |
| | | H | 24 | 1.22 | 0.00 | 0.00 | 0.00 | 1.22 | -0.22 | |
| Mg$_2$PtH$_6$ | 0.29 | Mg | 8 | 0.37 | 5.95 | 0.00 | 0.00 | 6.32 | 1.68 | [This] |
| | | Pt | 4 | 1.05 | 2.13 | 8.94 | 0.00 | 12.12 | -2.12 | |
| | | H | 24 | 1.21 | 0.00 | 0.00 | 0.00 | 1.21 | -0.21 | |

Magnesium (Mg) atoms carry a substantial positive Mulliken charge in all four compounds (approximately +1.52 to +1.83 e), indicating that Mg donates a significant portion of its electron density and behaves close to Mg$^{2+}$. The transition metal atoms (Rh, Pd, Ir, Pt) exhibit large negative Mulliken charges (−1.70 to −2.48 e), acting as electron-rich centers that accept charge from the surrounding Mg and H atoms. Their high d-orbital populations reflect significant occupation of metal d-states in the valence band. Hydrogen atoms display small negative Mulliken charges (around −0.20 to −0.23 e per H), suggesting that hydrogen primarily behaves as hydridic H⁻, but



with a noticeable covalent contribution. The Mulliken charge on hydrogen is far from a full −1 e, indicating that electron density is shared with both Mg and the transition metal. Among the transition metals, Pd and Pt exhibit slightly more negative Mulliken charges than Rh and Ir, while the Mg charges shift in the opposite direction to maintain overall charge balance. This trend reflects differences in electronegativity and bonding strength among Mg, the transition metal, and H, influencing the degree of electron transfer and covalency in the compounds.

**Table 13. Hirshfeld charge analysis for $Mg_2TmH_6$.**

| Compounds | All bands spilling Parameter (%) | Species | No. of Ions | Hirshfeld Charge (e) | Ref. |
|---|---|---|---|---|---|
| $Mg_2RhH_6$ | 0.86 | Mg | 8 | 0.27 | [This] |
|  |  | Rh | 4 | 0.01 |  |
|  |  | H | 24 | -0.09 |  |
| $Mg_2PdH_6$ | 0.74 | Mg | 8 | 0.19 | [This] |
|  |  | Pd | 4 | 0.29 |  |
|  |  | H | 24 | -0.11 |  |
| $Mg_2IrH_6$ | 0.83 | Mg | 8 | 0.30 | [This] |
|  |  | Ir | 4 | -0.17 |  |
|  |  | H | 24 | -0.07 |  |
| $Mg_2PtH_6$ | 0.76 | Mg | 8 | 0.23 | [This] |
|  |  | Pt | 4 | 0.07 |  |
|  |  | H | 24 | -0.09 |  |

Hirshfeld charges offer an alternative method for partitioning the electron density among atoms, based on a comparison of the molecular electron density with a superposition of spherical atomic densities. The column "All bands spilling parameter (%)" quantifies the fraction of charge not captured by this partitioning; values below 1% indicate that the analysis is numerically stable and reliable.

In the Hirshfeld analysis, the magnitude of the atomic charges is smaller than in the Mulliken scheme. Magnesium atoms carry modest positive charges (+0.19 to +0.30 e), confirming that Mg acts as an electron donor, but indicating less charge transfer than suggested by Mulliken analysis. The transition metals are near-neutral, with small positive or slightly negative Hirshfeld charges (Rh = +0.01 e, Pd = +0.29 e, Ir = −0.17 e, Pt = +0.07 e), reflecting a more even distribution of electron density among Mg, Tm, and H. Hydrogen atoms maintain small negative charges (−0.07



to −0.11 e), consistent with hydridic character, but again with reduced magnitude relative to Mulliken results.

Comparison of Mulliken and Hirshfeld analyses shows that, despite differences in numerical values, both methods yield a consistent qualitative picture of bonding. Significant charge transfer occurs from Mg to the Tm, $H_6$ framework, indicating that Mg behaves as a cationic sublattice. Hydrogen atoms carry small negative charges, supporting their role as hydrides and their participation in both Tm–H and Mg–H interactions. The transition-metal $d$-states are heavily involved in bonding, accommodating substantial electron density and mediating covalent interactions within the $TmH_6$ octahedra. Overall, the $Mg_2TmH_6$ compounds can be described as mixed ionic–covalent hydrides, with strongly electropositive Mg, slightly hydridic H, and electron-rich transition metals that act as covalently bonded centers.

### 3.8. Superconducting Properties

The superconducting behavior of a material is strongly influenced by its electronic structure, electron–phonon interactions, and electronic correlations. Key parameters such as the electron–phonon coupling constant, the Coulomb pseudopotential ($\mu^*$), and the density of states at the Fermi level determine both the onset and strength of superconductivity.

The repulsive Coulomb pseudopotential, $\mu^*$, can be determined from the TDOS at the Fermi level. The computed values of $\mu^*$ are listed in Table 14. We have calculated $\mu^*$ using the following equation [77]:

$$\mu^* = \frac{0.26 N(E_F)}{1 + N(E_F)} \tag{18}$$

Typical values of the Coulomb pseudopotential $\mu^*$ lie in the range 0.10–0.20 [86–89]. The calculated $\mu^*$ for $Mg_2TmH_6$ is relatively high, indicating that electronic correlations in this compound are significant.

We have calculated the superconducting critical temperatures of $Mg_2TmH_6$ using the following well-known McMillan equation [78]:



$$T_C = \frac{\theta_D}{1.45} \exp\left[-\frac{1.04(1+\lambda_{ep})}{\lambda_{ep}-\mu^*(1+0.62\lambda_{ep})}\right] \tag{19}$$

For the present calculations, the electron–phonon coupling constants were estimated using the previously reported values [34], assuming a linear dependence on $N(E_F)$ expressed as $\lambda = N(E_F)V_{\text{e-ph}}$, where $V_{\text{e-ph}}$ is the electron–phonon interaction strength, which is considered the same for this work and the previous one [34]. This is an approximation which has been found to be largely valid for a number of phonon mediated superconductors [73]. The superconducting transition temperature ($T_C$) is then evaluated using the computed $N(E_F)$, $\mu^*$, $\lambda$, and $\theta_D$ with the help of Eq. (19). The calculated values of $T_C$ are listed together with the relevant parameters in Table 14 and compared with the previously reported values [34]. In our calculations, the highest value of $T_C$ is obtained for $Mg_2PdH_6$, whereas the previous study [34] reported $Mg_2PtH_6$ as having the highest transition temperature. This discrepancy can be attributed to differences in the calculated $N(E_F)$, phonon spectra, and electron–phonon matrix elements, underscoring the sensitivity of superconducting properties to detailed electronic-structure and lattice-dynamical treatments. Nevertheless, the overall trend across the $Mg_2TmH_6$ family remains consistent between the two studies.

Table 14. TDOS at the Fermi level ($N(E_F)$ in states/eV), repulsive Coulomb pseudopotential ($\mu^*$), electron-phonon coupling constant ($\lambda$), and superconducting transition temperature ($T_C$ in K) of $Mg_2TmH_6$ ($Tm$ = Rh, Pd, Ir, Pt) in comparison with the previous value.

| Compounds | $N(E_F)$ | $\mu^*$ | $\lambda$ | $T_C$ | $N(E_F)$ | $\mu^*$ | $\lambda$ | $T_C$ |
|---|---|---|---|---|---|---|---|---|
| | [This] | | | | [34] | | | |
| $Mg_2RhH_6$ | 1.93 | 0.171 | 1.19 | 25.07 | 2.1 | 0.14 | 1.3 | 48.50 |
| $Mg_2PdH_6$ | 1.43 | 0.153 | 1.75 | 44.53 | 0.9 | 0.14 | 1.1 | 66.50 |
| $Mg_2IrH_6$ | 1.54 | 0.158 | 1.24 | 26.34 | 2.6 | 0.14 | 2.1 | 77.00 |
| $Mg_2PtH_6$ | 1.19 | 0.141 | 1.67 | 38.01 | 1.0 | 0.14 | 1.4 | 80.04 |

### 3.9. Optical Properties

The optical properties of metals are determined by their interaction with electromagnetic radiation, which is closely linked to their electronic band structure. Free electrons in the metal can respond to incident light, giving rise to characteristic phenomena such as reflection, absorption, and plasmonic resonances. The behavior of these electrons is often described by the Drude model [79,80], which accounts for the collective motion of conduction electrons and explains the high reflectivity of metals at visible and infrared frequencies. In evaluating the optical response of metallic Mg$_2$TmH$_6$, a screened plasma energy of 10 eV and a Drude damping value of 0.1 eV have employed. Optical responses also depend on interband transitions, where electrons are excited from filled to empty states, influencing absorption spectra at higher energies.

The dielectric functions [see Fig. 15(a)] of Mg$_2$RhH$_6$, Mg$_2$PdH$_6$, Mg$_2$IrH$_6$, and Mg$_2$PtH$_6$ show similar overall behavior with compound dependent differences in magnitude and peak positions. In all cases, the real part of the dielectric constant is negative at low frequencies, indicating Drude-like behavior. The imaginary part of the dielectric function, $\varepsilon_2(\omega)$, represents the absorptive component of the optical response and accounts for the energy dissipation of electromagnetic radiation within the material. In metals, the low-frequency region is predominantly governed by intraband (free-carrier) transitions, which are well described by the Drude model. As a result, $\varepsilon_2(\omega)$ exhibits a large magnitude at low frequencies. With increasing photon energy, the intraband contribution gradually diminishes in the visible region. At higher energies, particularly in the ultraviolet region, $\varepsilon_2(\omega)$ increases again owing to interband electronic transitions. Pronounced peaks in 8 – 10 eV energy region confirm significant optical absorption and high electronic polarizability. Differences among the compounds appear mainly in the intensity and energy shift of these oscillations and absorption peaks, which are controlled by changes in the electronic structure and the contribution of the transition metal $d$ states. Mg$_2$IrH$_6$ and Mg$_2$PtH$_6$ show more noticeable shifts due to their heavier Ir and Pt elements, respectively. At higher photon energies, both the real and imaginary parts gradually approach small values for all compounds, indicating reduced dispersion and the weakening of optical transitions beyond the main absorption region.

Reflectivity spectra [Fig. 15(b)] of Mg$_2$TmH$_6$ exhibit a common metal-like response at low photon energies. All compounds show very high reflectivity (close to unity) in the low-energy region, spanning the infrared and extending into the visible range. A pronounced decrease occurs near the first major absorption edge, where strong interband transitions set in. In the mid-energy range (7 – 17 eV), the reflectivity recovers, displaying a series of moderate peaks and oscillations that follow the detailed structure of the dielectric response, indicating sustained light–matter interaction. Compared with the other compounds Mg$_2$PdH$_6$ tends to maintain slightly higher reflectivity over portions of the mid-energy region, while Mg$_2$PtH$_6$ exhibits comparable but occasionally slightly lower values. Mg$_2$IrH$_6$ generally lies between Mg$_2$PdH$_6$ and Mg$_2$PtH$_6$ within the same energy range. At higher photon energies, fine oscillations persist for all compounds, which can be attributed to overlapping interband transition processes.

The refractive index and extinction coefficient spectra of studied hydrides [Fig. 15(c)] exhibit similar trends with compound dependent variations in peak intensity and position. All compounds show a large static refractive index at low photon energies, indicating strong interaction with low energy light and a pronounced reduction of wave velocity inside the material. With increasing photon energy, the refractive index decreases rapidly and then approaches an almost constant value in the high energy region, typically beyond about 10 eV. The extinction coefficient displays pronounced maxima in the low to mid energy range, aligning with the energies where absorption is strongest and where the imaginary part of the dielectric function is large. These peaks reflect strong attenuation of electromagnetic waves due to intense interband transitions, while at higher photon energies the extinction coefficient becomes smaller, consistent with reduced optical absorption and weaker dispersion.

The absorption coefficient quantifies the rate at which optical intensity decays inside a solid, with the penetration depth approximately given by $1/\alpha$. A pronounced increase in $\alpha$ occurs when incident photons efficiently promote electrons through available interband transitions, a behavior commonly observed in the ultraviolet region of transition-metal compounds. $Mg_2RhH_6$ and $Mg_2PdH_6$ exhibit strong and broad UV absorption across the mid-energy range, indicating short penetration depths and efficient interband excitation over wide spectral intervals (Fig. 15(d)). Among the series, $Mg_2IrH_6$ shows the highest absorption intensity in the mid-energy region, reflecting particularly strong transition activity, while $Mg_2PtH_6$ maintains high absorption throughout the same UV window with slightly reduced peak intensity compared to $Mg_2IrH_6$. Because the absorption coefficient remains high in the ultraviolet region, these hydrides may be considered promising for UV optoelectronic and UV-shielding applications [81].

Fig. 15(e) depicts the optical conductivity spectra of $Mg_2TmH_6$ which exhibit a large value at zero or very low photon energies, followed by a rapid decrease in the low-energy region. This is then succeeded by a sharp increase in the visible to near-ultraviolet region, where pronounced maxima are observed. These main conductivity peaks coincide with the strongest absorption features, indicating efficient photoexcited carrier generation driven by intense interband transitions. At higher photon energies, the optical conductivity gradually decreases for all compounds, reflecting the reduced density of available optical transitions.

The energy loss function is derived from the complex dielectric response through $\text{Im}[-1/\varepsilon(\omega)]$ and it describes screened electronic excitations in a solid. A dominant maximum in this spectrum corresponds to the bulk plasmon resonance and its position gives the bulk plasma frequency [82]. For $Mg_2RhH_6$ a clear plasmon peak appears at 19.54 eV, indicating a well-defined collective electron oscillation with limited damping. $Mg_2PdH_6$ shows a pronounced peak at 17.82 eV, slightly shifted due to differences in effective electron density. $Mg_2IrH_6$ exhibits a strong peak at 19.28 eV, consistent with a robust plasmon mode. $Mg_2PtH_6$ presents an intense peak at 18.59 eV, confirming a characteristic plasma frequency similar in strength but shifted in energy relative to the other hydrides.

Overall, $Mg_2TmH_6$ (*Tm* = Rh, Pd, Ir, Pt) show strong interband-driven optical response with high UV absorption and associated optical conductivity, which supports potential UV optoelectronic and UV photodetector type uses [81]. Their high low-energy reflectivity also suggests utility in reflective optical coatings and mirror type components [83].

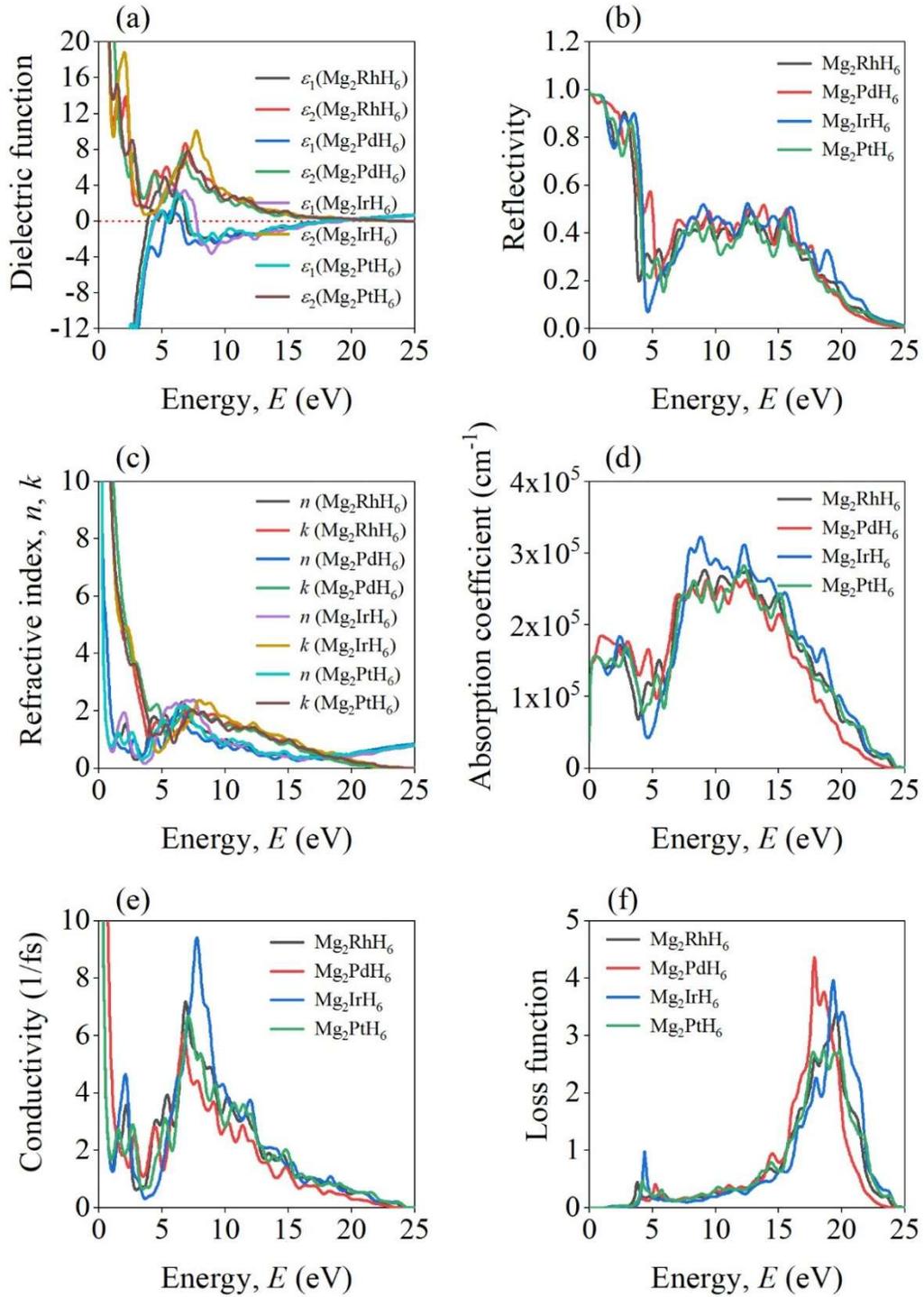

**Fig. 15. Optical properties of Mg$_2$*Tm*H$_6$ (*Tm* = Rh, Pd, Ir, Pt).**

## 4. Conclusions

In this study, we have systematically investigated the structural, hydrogen storage, electronic, elastic, mechanical, thermophysical, superconducting, and optical properties of $Mg_2TmH_6$ ($Tm$ = Rh, Pd, Ir, Pt) hydrides. These compounds are thermodynamically, mechanically and dynamically stable and exhibit good hydrogen storage capacities (2.42–3.84 wt%). All hydrides are ductile and elastically anisotropic, with weak metallic character and strong directional bonding, reflected in the elastic hierarchy $C_{11} > C_{12} > C_{44}$ and the moduli trend $Y > B > G$. Among the compounds, $Mg_2PtH_6$ shows the largest bulk modulus, indicating the lowest compressibility, while $Mg_2IrH_6$ has the highest Young's modulus, reflecting the greatest resistance to elastic deformation, along with the highest machinability index and excellent dry lubricity. Hardness values are low to moderate (within 1–5 GPa). Moderate hardness, ductility and machinability index comparable to MAX phase compounds [84-86] imply that $Mg_2TmH_6$ ($Tm$ = Rh, Pd, Ir, Pt) materials hold promise in structural engineering sector. The electronic structures yield low total density of states at the Fermi level, resulting in predicted superconducting transition temperatures between 25 and 44 K. Thermophysical parameters including the Grüneisen and Kleinman parameters, Debye and melting temperatures, thermal-expansion coefficient, and lattice/minimum thermal conductivities provide a comprehensive view of their stability and heat transport. Low minimum thermal conductivity and moderate melting temperature of $Mg_2IrH_6$ and $Mg_2PtH_6$ suggest that these two compounds have potential to be used as thermal barrier coating [87]. Optical analyses disclose high reflectivity in the infrared and visible regions, with a noticeable decline in the ultraviolet range. Strong UV absorption and significant optical conductivity designate their suitability for UV optoelectronic devices, photodetectors, and reflective optical coatings.

Overall, $Mg_2TmH_6$ hydrides primarily offer favorable hydrogen storage capacity while also providing mechanical robustness, high-temperature superconductivity, and multifunctional optical properties. These attributes make them promising candidates for superconducting, energy storage and advanced optoelectronic applications.

## Acknowledgements

S. H. N. acknowledges the research grant (1151/5/52/RU/Science-07/19-20) from the Faculty of Science, University of Rajshahi, Bangladesh, which partly supported this work. M. A. A.

acknowledges the financial support from the Science and Technology Fellowship Trust, Ministry of Science and Technology, Bangladesh for his Ph.D. research.

**Data availability**

The data sets generated and/or analyzed in this study are available from the corresponding author on reasonable request.

**Declaration of interest**

The authors declare that they have no known competing financial interests or personal relationships that could have appeared to influence the work reported in this paper.

**CRediT author statement**

**M. A. Alam:** Methodology, Software, Investigation, Resources, Visualization, Writing- Original draft, **M. A. H. Shah:** Visualization, Resources, Formal analysis, Writing- Reviewing and Editing, **F. Parvin:** Supervision, Validation, Writing-Reviewing and Editing. **S. H. Naqib:** Conceptualization, Supervision, Validation, Administration, Writing- Reviewing and Editing.